\documentclass[aps,pra,twocolumn,10pt,longbibliography,nobibnotes]{revtex4-2}
\usepackage{amsmath,amssymb,graphicx,mathtools,dsfont,courier}
\usepackage[T1]{fontenc}

\begin{document}

\newcommand{\bra}[1]    {\langle #1|}
\newcommand{\ket}[1]    {|#1 \rangle}
\newcommand{\ketbra}[2]{|#1\rangle\!\langle#2|}
\newcommand{\braket}[2]{\langle#1|#2\rangle}
\newcommand{\tr}[1]    {{\rm Tr}[ #1 ]}
\newcommand{\trr}[2]    {{\rm Tr}[ #1 ]_{\overline{#2}}}
\newcommand{\titr}[1]    {\widetilde{{\rm Tr}}\left[ #1 \right]}
\newcommand{\av}[1]    {\langle #1 \rangle}
\newcommand{\modsq}[1]    {| #1 |^2}
\newcommand{\0}    {\ket{\vec 0}}
\newcommand{\1}    {\ket{\vec 1}}
\newcommand{\dt}    {\delta\theta}
\newcommand{\I}    {\mathcal  I_{an}^{(q)}}
\newcommand{\Iq}    {\mathcal  I_{q}}
\newcommand{\Id}[1]    {\mathcal  I_q\left[#1\right]}
\newcommand{\Ic}[1]    {\mathcal  I_{cl}\left[#1\right]}
\newcommand{\Icl}    {\mathcal  I_{cl}}
\newcommand{\C}    {\hat{\mathcal C}(t)}
\newcommand{\Cd}    {\hat{\mathcal C}^\dagger(t)}
\newcommand{\re}[1]    {\mathrm{Re}\left[#1\right]}
\newcommand{\im}[1]    {\texttt{Im}\left[#1\right]}

\title{Metrology using atoms in an array of double-well potentials}
\author{Danish Ali Hamza and Jan Chwede\'nczuk}
\affiliation{Faculty of Physics, University of Warsaw, ulica Pasteura 5, 02-093 Warszawa, Poland}

\begin{abstract}
  Quantum effects, such as entanglement, Einstein-Podolsky-Rosen steering, and Bell correlations,
  can enhance metrological sensitivity beyond the standard quantum limit. These correlations are typically generated through
  interactions between atoms or molecules, or during the passage of a laser pulse through a birefringent crystal. 
  Here, we consider an alternative method of generating scalable, many-body entangled states, 
  and demonstrate their usability for quantum-enhanced metrology. Our setup is a one-dimensional (1D) array of double-well potentials holding independent and uncorrelated Bose-Einstein condensates. 
  The beam-splitting transformation mixes the signal between adjacent wells and yields a strongly entangled state through a many-body equivalent of the Hong-Ou-Mandel effect. We demonstrate this entanglement
  can improve the sensitivity of quantum sensors. In our analysis, we account for the effects of atomic fluctuations and identify the optimal measurement that saturates the quantum Cramer-Rao bound.
\end{abstract}

\maketitle

\section{Introduction.}

``Be precise. A lack of precision is dangerous when the margin of error is small''---Donald Rumsfeld, an American buisnessman and politician, once said. Same in physics---precision of measurements
can be crucial. A measurement is only accurate when the sensor is highly sensitive to small changes in the observed system~\cite{braunstein1994statistical}. When restricting to classical phenomena only,
the sensitivity $\Delta\theta$ of a device that measures some parameter $\theta$ and consists of $N$ detecting particles is restricted 
by the  Standard Quantum Limit (SQL)~\cite{holevo2011probabilistic,giovannetti2004quantum}
\begin{align}\label{eq.sql}
  \Delta\theta\geqslant\frac1{\sqrt N}.
\end{align}
If $N$ can be arbitrarily large, then so can the sensitivity. However, the number of resources is often limited.
Gravitational wave detectors cannot use lasers of infinite
intensity because the sensitivity would quickly degrade due to recoil or heating of the mirrors~\cite{abbott2016observation}. 
When measuring subtle variations in gravitational acceleration or magnetic field fluctuations, 
one should also use small and appropriately subtle atomic, photonic, molecular or solid-state 
sensors~\cite{ferrari,poli,PhysRevA.91.033629,PhysRevA.74.023615,PhysRevA.65.033608,kasevich1992measurement,doi:10.1126/science.1135459,doi:10.1126/science.1135459,fattori,PhysRevLett.120.013401}.

Quantum effects come into play at this moment. In principle, one can achieve a balance between a finite $N$ and high sensitivity by utilizing one from the triad of quantum resources:
entanglement~\cite{giovannetti2004quantum,dunn2,pezze2009entanglement}, Einstein-Podolsky-Rosen steering~\cite{yadin2021metrological}, or Bell correlations~\cite{PhysRevLett.126.210506,PhysRevA.99.040101}. 
The ultimate bound is the Heisenberg limit (HL)~\cite{pezze2009entanglement,PhysRevA.68.025602}
\begin{align}\label{eq.hl.intro}
  \Delta\theta\geqslant\frac1{N}.
\end{align}
To reach the HL, the most subtly correlated quantum states must be used. However, due to the immediate decoherence, it is rather elusive~\cite{dobrz}. Therefore whenever an improvement to the SQL
is demonstrated, i.e., $\Delta\theta<\frac1{\sqrt N}$, it is considered a major achievement. Entanglement in multi-qubit systems is often generated by 
spin squeezing the sample~\cite{schumm2005matter,jo2007long,bohi2009coherent,esteve2008squeezing,huang2008optimized,leroux2010orientation,chen2011conditional,berrada2013integrated}. 
Another
method is to use two-body collisions of atoms to create an intense source of entangled pairs~\cite{twin_beam,cauchy_paris,collision_paris,twin_paris,ketterle,truscott,shin2019bell,PhysRevLett.112.155304,doi:10.1126/science.1208798} in analogy to the well-known optical parametric down-conversion process~\cite{pdc1,pdc2}.

\begin{figure}[t!]
    \centering
    \includegraphics[width=.9\linewidth]{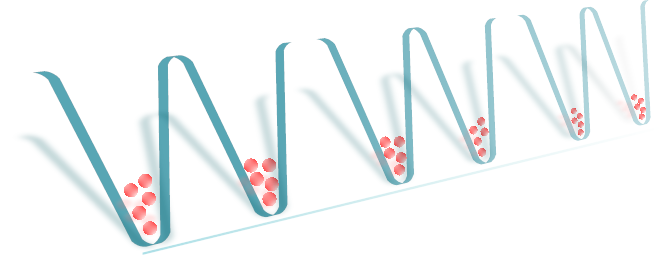}
    \caption{This is an illustration of the interferometric setup of interest in this work. It shows a collection of $M$ double wells, each of which is filled with an ultra-cold bosonic gas.}
    \label{fig.array}
\end{figure}

Sensitivity can also be improved by increasing the signal dependent on $\theta$. 
In gravitational-wave detection, the signal is amplified using the large, 4-kilometer-long arms of the Michelson interferometer~\cite{abbott2016observation}.
For measurements of a local gravitational acceleration, the two atomic beam trajectories are separated as far as matter-wave coherence constraints allow~\cite{A_Peters_2001,Tino_2021,peters1999measurement,peters2001high,weiss1994precision,Altin_2013}.
In this vein is the recent series of experiments 
involving a one-dimensional array of double-well (DW) potentials that trap ultra-cold bosonic atoms, schematically shown in Fig.~\ref{fig.array}. 
The array is lengthy; therefore, if each DW acts as an interferometer measuring $\theta$, the overall precision can be high. Long-range coherence has been demonstrated in this system through 
the observation of long-lived Bloch oscillations~\cite{masi2021spatial}. Recently, this system was used for the first time to precisely measure gravitational acceleration~\cite{petrucciani2025mach}.

This manuscript aims to explain how the sensitivity of the multi-DW interferometer increases due to the interplay between its large size and quantum correlations.
First, we introduce the model in Section~\ref{sec.model}. Next, Section~\ref{sec.tool} briefly discusses the main tool that determines sensitivity: the Quantum Fisher Information (QFI).
In Section~\ref{sec.ult} we derive the ultimate bounds (i.e., the SQL and the HL) for this system. This part of the mansuscript concludes with the discussion of mode mixing, which is equivalent to
an inter-DW beam-splitter, see Section~\ref{sec.mix}. This mixing creates a nontrivial multimode state across the array by correlating the modes.
The next section discusses the impact of this mixing on sensitivity. We demonstrate that, even when an interferometer is supplied with a product of ``semiclassical'' atomic states, 
bosonic interference at the beam splitter can enhance sensitivity (see Section~\ref{sec.css}). This is a many-body analog of the Hong-Ou-Mandel (HOM) effect~\cite{hong}.
Next, in Section~\ref{sec.oat}, we compare this result to the case in which the interferometer is fed by a collection of entangled states obtained through the one-axis twisting (OAT) procedure.
It is a well-known method for producing highly entangled and even Bell-correlated states, which are useful for quantum metrology~\cite{esteve2008squeezing,plodzien2022one}. 
This is why we are interested in this technique.
In Section~\ref{sec.fluct} we illustrate the impact of shot-to-shot atom number fluctuations on the sensitivity.
The Appendix follows the conclusions. We moved some of the details of the analytical calculations there for the sake of clarity in the manuscript.

\section{Model}

In this section, we will discuss the basic building blocks of the multi-DW interferometer: state, sensitivity, and allowed transformations.

\subsection{The state}\label{sec.model}

The system under consideration is comprised of $N$ bosonic particles of a single species, which are distributed across $M$ double-well potentials, forming a 1D array, as illustrated in Fig. ~\ref{fig.array}. 
\begin{figure}
  \centering
  \includegraphics[width=\linewidth]{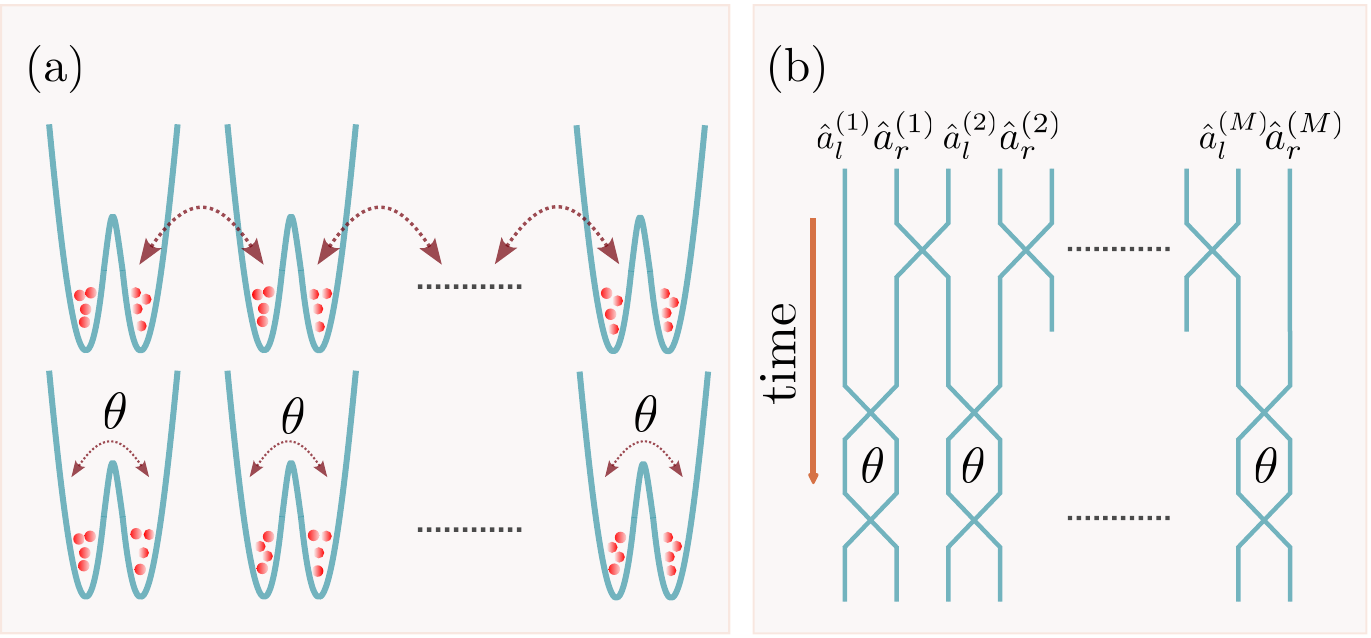}
  \caption{(a): The illustration of the interferometric procedure.  The top row shows the mixing of modes of adjacent double wells. The bottom one depicts the imprinting of parameter $\theta$ using
    the collection of $M$ independent Mach-Zehnder interferometers. (b): Schematic representation showing the two steps brought together to form a time sequence.}
  \label{fig.scheme}
\end{figure}
The full Hilbert space $\mathcal H$ is spanned by Fock states
\begin{align}\label{eq.vecn}
  \ket{\vec n}=\ket{n_l^{(1)},n_r^{(1)}\ldots n_l^{(M)},n_r^{(M)}},
\end{align}
where $n_{l/r}^{(i)}$ is the number of bosons in the left/right mode of the $i$-th well with the constraint
\begin{align}
  \sum_{i=1}^M\left(n_l^{(i)}+n_r^{(i)}\right)=N.
\end{align}
In this configuration, any quantum state can be represented as
\begin{align}
  \hat\varrho=\sum_{\vec n,\vec n'}\varrho_{ij}\ketbra{\vec n}{\vec n'}.
\end{align}
Nevertheless, it is challenging to work with states of this form, as the dimension of $\mathcal{H}$ is too large to address the problem, even when $N$ and $M$ are moderate. 
The focus henceforth will be restricted to an experimentally substantiated case. Initially, the double-wells form separate subsystems, each occupied by $n$ bosons, so that $n\,M=N$. This vastly simplifies the analysis as
now any quantum state can be expressed as follows
\begin{align}\label{eq.m.sep}
  \hat\varrho=\sum_mp_m\hat\varrho_1^{(m)}\ldots\hat\varrho_M^{(m)},
\end{align}
where $p_m\geqslant9$ and $\sum_mp_m=1$, while each of the ``local'' density operators can be expressed by means of vectors $\ket{n_r,n-n_r}$. 
While the focus will be on pure input states in establishing upper bounds for the sensitivity, mixed state will be considered in the context of potential experimental imperfections. 

\subsection{The tool: Quantum Fisher Information}\label{sec.tool}

Inspired by a recent experimental work~\cite{petrucciani2025mach} we consider $M$ Mach-Zehnder interferometers, feeding independently each DW with the  information about an unknown parameter $\theta$.
This can be expressed in terms of a unitary transformation of an input state  $\hat\varrho$, as follows:
\begin{align}\label{eq.dens.th}
  \hat\varrho(\theta)=e^{-i\theta\hat J_y}\hat\varrho\, e^{i\theta\hat J_y},
\end{align}
where the collective operator is
\begin{align}\label{eq.gen}
  \hat J_y\equiv \sum_{i=1}^M\hat J_y^{(i)}.
\end{align}
and the individual single-DW operators are defined as follows
\begin{align}\label{eq.def.jy}
  \hat J_y^{(i)}=\frac1{2i}\left(\hat a_r^{(i)\dagger}\hat a_l^{(i)}-\hat a_l^{(i)\dagger}\hat a_r^{(i)}\right).
\end{align}
The operators $\hat a_{l/r}^{(i)}$ annihilate a boson in the left/right well of the $i$-th DW. 
A transformation like in Eq.~\eqref{eq.dens.th} can be triggered by a linear external potential acting on the system, 
so the scenario here could be relevant to precise gravitometry of the local gravitational acceleration  $g$ \cite{petrucciani2025mach}. 

Once the output state is generated, some measurements yield the estimated parameter with an uncertainity $\pm\Delta\theta$. Its minimal value is given
by the quantum Cramer-Rao lower bound (QCRB)
\begin{align}\label{eq.qcrb}
  \Delta\theta\geqslant\frac1{\sqrt{\Id{\hat\varrho(\theta)}}},
\end{align}
where $\Id{\hat\varrho(\theta)}$ is the quantum Fisher information (QFI) that reads~\cite{braunstein1994statistical}
\begin{align}\label{eq.qfi.gen}
  \Id{\hat\varrho(\theta)}=2\sum_{i,j}\frac1{p_i+p_j}\modsq{\bra{\psi_i}\dot{\hat\varrho}(\theta)\ket{\psi_j}}.
\end{align}
Here, the dot denotes the derivative of the density matrix from Eq.~\eqref{eq.dens.th} over $\theta$ and $\ket{\psi_{i/j}}$ are the eigen-vectors of the density matrix $\hat\varrho(\theta)$ with the
corresponding eigen-values $p_{i/j}$. The double sum runs through all the non-trivial part of the spectrum, i.e., whenever $p_i+p_j\neq0$.

The analytical evaluation of the QFI, necessary to establish the QCRB from Eq.~\eqref{eq.qcrb}, 
is usually impossible as the analytical diagonalization of the 
$\theta$-dependent density matrix is hardly ever possible. However, a major simplification comes from the following inequality~\cite{braunstein1994statistical}
\begin{align}\label{eq.qfi.pure}
  \Id{\hat\varrho(\theta)}\leqslant4\Delta^2\hat J_y\equiv4\left(\av{\hat J_y^2}-\av{\hat J_y}^2\right)\leqslant4\av{\hat J_y^2},
\end{align}
where the averages are calculated using $\av{\cdot}=\tr{\hat\varrho(\theta)\cdot}$. This inequality is saturated if the density matrix $\hat\varrho(\theta)$ represents a pure state. 
Consequently, in order to establish the upper bounds for the sensitivity, it is essential to identify the optimal pure states. In the following section, we will proceed with a detailed analysis of this concept.

\subsection{Ultimate scalings}\label{sec.ult}

We begin by establishing a shot-noise  limit for this system. This will allow us to track any quantum enhancements due to the entanglement of the input state of the collective Mach-Zehnder interferometer.
A separable state will be one that does not exhibit quantum correlations among the DWs, i.e., such as in Eq.~\eqref{eq.m.sep}. Moreover, the two-mode $n$-boson density matrix of every DW, $\hat\varrho_i$,
must represent a ``classical'' state. The basic building block of such a $\hat\varrho_i$ is called a coherent spin state (CSS), which reads
\begin{align}\label{eq.css}
  \ket{\theta_i,\varphi_i}=\frac1{\sqrt{n!}}\left(\hat b^{(i)\dagger}\right)^n\ket{0}.
\end{align}
It is simply a Fock state, obtained with an $n$-fold action of a creation operator 
\begin{align}\label{eq.css.cr}
  \hat b^{(i)\dagger}=\hat a^{(i)\dagger}_r\cos\theta_i+\hat a^{(i)\dagger}_l\sin\theta_i e^{i\varphi_i}
\end{align}
with $\theta_i\in[0,\pi[$ and $\varphi_i\in[0,2\pi[$, on the two-mode vacuum $\ket0$. States~\eqref{eq.css} can be used to construct any separable single-DW state in the form
\begin{align}\label{eq.css.mix}
  \hat\varrho_i=\int_0^{\pi}\!d\theta_i\int_0^{2\pi}\!d\varphi_i\,\mathcal P(\theta_i,\varphi_i)\ketbra{\theta_i,\varphi_i}{\theta_i,\varphi_i},
\end{align}
where $\mathcal P$ is any probability distribution of variables $\theta_i$ and $\varphi_i$. Subsequently, a separable state of $M$ double-wells is constructed as a product
\begin{align}\label{eq.css.sep}
  \hat\varrho_{sep}=\bigotimes_{i=1}^M\hat\varrho_i.
\end{align}
Finally, one can consider a statistical mixture of such product states, as in Eq.~\eqref{eq.m.sep}, where $m$ labels the different probability distributions $\mathcal P^{(m)}(\theta_i,\varphi_i)$.
In this way, the most general separable state is constructed.

This state sets a reference value of the sensitivity as it gives
\begin{align}\label{eq.snl}
  \Delta\theta\geqslant\frac1{\sqrt{M}}\frac1{\sqrt{n}}=\frac1{\sqrt N}, 
\end{align}
i.e., a shot-noise scaling with the total number of resources, $N$, or a ``double'' shot-noise scaling with the number of DWs, $M$, and the number of atoms per DW, $n$. 

Values of $\Delta\theta$ that are smaller than the SQL of Eq.~\eqref{eq.snl} imply the presence of some form of entanglement in the system. 
For example, $\hat\varrho$ might still be expressed as in Eq.~\eqref{eq.m.sep} but at least one of the local states could be non-separable, which means it could not be expressed as in Eq.~\eqref{eq.css.mix}.
The limiting value of the sensitivity for such a case is obtained when the interferometer is fed with a product of $M$ NOON states
\begin{align}\label{eq.noon.local}
  \ket{\psi^{(i)}}=\frac1{\sqrt2}(\ket{n,0}_y^{(i)}+\ket{0,n}_y^{(i)}),
\end{align}
giving the Heisenberg scaling with $n$, namely
\begin{align}\label{eq.hl.local}
  \Delta\theta=\frac1{\sqrt{M}}\frac1{n}. 
\end{align}
Here, the two components of the superposition~\eqref{eq.noon.local} are the eigenstates of $\hat J_y^{(i)}$ with the maximal/minimal eigenvalue, namely
\begin{subequations}
  \begin{align}
    &\hat J_y^{(i)}\ket{n,0}_y^{(i)}=\frac n2\ket{n,0}_y^{(i)},\\
    &\hat J_y^{(i)}\ket{0,n}_y^{(i)}=-\frac n2\ket{0,n}_y^{(i)}.
  \end{align}
\end{subequations}
The ``intra-entanglement''  depicts situations when the state cannot be represented in the form of Eq.~\eqref{eq.m.sep}. The DWs become entangled and a global NOON state
\begin{align}\label{eq.noon.global}
  \ket{\psi}=\frac1{\sqrt2}\left(\bigotimes_{i=1}^M\ket{n,0}_y^{(i)}+\bigotimes_{i=1}^M\ket{0,n}_y^{(i)}\right),
\end{align}
gives the global Heisenberg scaling with the total number of resources, i.e., 
\begin{align}\label{eq.hl}
  \Delta\theta=\frac1{M}\frac1{n}=\frac1N. 
\end{align}
The values of $\Delta\theta$ in the range between Eqs~\eqref{eq.snl} and~\eqref{eq.hl} are the ``playground'' for quantum-enhanced metrology with a collection of DWs. 
The pair of states, Eqs~\eqref{eq.noon.local}
and~\eqref{eq.noon.global}, that saturate~\eqref{eq.hl.local} and ~\eqref{eq.hl} respectively, set a reference, 
but should not be pursued as an achievable goal. Despite impressive progress made in trapped atom interferometry, 
including the multi-DW setup~\cite{masi2021spatial,petrucciani2025mach}, the entangled systems that can be generated at present (or in the nearby future)
are a far cry from those NOON states.
Therefore in the next section we propose a modificiation of the MZI sequence of Eq.~\eqref{eq.dens.th} that still beats the SQL but using CSSs at the input.

\subsection{Mode mixing}\label{sec.mix}

Consider an additional operation, equivalent to a beam splitter, that couples the right mode of the $i$-th DW with the left mode of the $(i+1)$, as shown schematically in Fig.~\ref{fig.scheme}(a). 
It is generated by 
\begin{align}\label{eq.mix}
  \hat S_x^{(i)}=\frac1{2}\left(\hat a_r^{(i)\dagger}\hat a_l^{(i+1)}+\hat a_l^{(i+1)\dagger}\hat a_r^{(i)}\right).
\end{align}
The collective beam-splitter operator is thus
\begin{align}
  \hat S_x=\sum_{i=1}^{M-1}\hat S_x^{(i)}
\end{align}
(the truncation of the sum at $i=M-1$ is a consequence of open boundary conditions). We assume that this transformation precedes the MZI interferometers and modifies the input state as follows:
\begin{align}\label{eq.dens.mix}
  \hat\varrho\ \ \longrightarrow\ \ e^{-i\frac\pi2\hat S_x}\hat\varrho\, e^{i\frac\pi2\hat S_x}.
\end{align}
Subsequently, this density matrix is fed into the collection of MZIs, see Eq.~\eqref{eq.dens.th}. 

For convenience, it is preferable to switch from the Schr\"odinger picture [see Eq.~\eqref{eq.dens.mix}] to the Heisenberg picture and evolve the generator of the MZI transformation, Eq.~\eqref{eq.gen}, namely
\begin{align}
  \hat S_y\equiv e^{i\frac\pi2 \hat S_x}\hat J_y e^{-i\frac\pi2 \hat S_x}
\end{align}
or equivalently
\begin{align}\label{eq.sy.ind}
  \hat S^{(i)}_y\equiv e^{i\frac\pi2 \hat S_x}\hat J^{(i)}_y e^{-i\frac\pi2 \hat S_x}.
\end{align}
By inspecting the expression for the mixing operator $\hat S_x$, Eq.~\eqref{eq.mix}, we obtain the pairs of left / right annihilation operators that couple to its closest neighbours
\begin{subequations}
  \begin{align}
    &\hat a_{(l),out}^{(i-1)}=\frac{\hat a_{l}^{(i-1)}-i \hat a_r^{(i)}}{\sqrt{2}},\\
    &\hat a_{(r),out}^{(i)}=\frac{\hat a^{(i)}_{r}-i \hat a_{l}^{(i-1)}}{\sqrt{2}}.
  \end{align}
\end{subequations}
Consequently, the internal operators labelled with $i\neq1,M$ (i.e., those that couple to the neighbours at both sides) take the following form:
\begin{align}
  \hat S^{(i)}_{y}=\frac12\left(\hat J^{(i)}_{y}+\hat Y^{(i)}_{y}+\hat R^{(i)}-\hat L^{(i+1)}\right).
\end{align}
where the three new operators are
\begin{subequations}
  \begin{align}
    &\hat L^{(i)}=\frac12\left(\hat a_l^{(i-1)\dagger}\hat a_l^{(i)}+\hat a_l^{(i-1)}\hat a_l^{(i)\dagger }\right) \\
    &\hat R^{(i)}=\frac12\left(\hat a_r^{(i-1)\dagger}\hat a_r^{(i)}+\hat a_r^{(i-1)}\hat a_r^{(i)\dagger}\right)\\
    &\hat Y^{(i)}=\frac{1}{2i}\left(\hat a_l^{(i-1)\dagger}\hat a_r^{(i+1)}-\hat a_r^{(i+1)\dagger}\hat a_l^{(i-1)}\right). 
  \end{align}
\end{subequations}
Finally, the extreme operators, i.e., for $i=1$ and $i=M$ transform into
\begin{subequations}
  \begin{align}
    &\hat S^{(1)}_{y}=\frac1{\sqrt2}\left(\hat J^{(1)}_{y}-\hat L^{(2)}\right),\\
    &\hat S^{(M)}_{y}=\frac1{\sqrt2}\left(\hat J^{(M)}_{y}+\hat R^{(M)}\right).
  \end{align}
\end{subequations}
We are now prepared to compute the sensitivity for the mixing / MZI protocol with some semi-classical input states. 
\section{Breaking the SQL}

Using the expressions from Eqs~\eqref{eq.qfi.pure} and~\eqref{eq.sy.ind} we obtain
\begin{align}\label{eq.qfim}
    \Iq=4\sum_{i=1}^M\Delta^2\hat S^{(i)}_{y}+8\sum_{i\neq j=1}^M\left(\av{\hat S^{(i)}_{y}\hat S^{(j)}_{y}}-\av{\hat S^{(i)}_{y}}\av{\hat S^{(j)}_{y}}\right)
\end{align}
which is used below to establish the QCRB for pure states.

\subsection{Coherent spin states}\label{sec.css}

First, we consider the CSS, see the definition in Eq.~\eqref{eq.css}, with $\theta_i=0$ for all $i$. This corresponds to
\begin{align}\label{eq.css.2}
  \ket{\psi}=\bigotimes_{i=1}^M\ket{0,n}_i.
\end{align}
Such a state is an idealization of a setup, where $M$ independent BECs are loaded into every second site of a multi-DW system. In contrast to a realistic scenario, it neglects the shot-to-shot
atom-number fluctuations, which will be discussed in the next section. Using Eq.~\eqref{eq.qfim} we obtain [see Appendix~\ref{app1} for details]
\begin{equation}\label{eq.scale.fock}
    \Iq=\frac 12M(n^2+2n).
\end{equation}
Though the sensitivity scales at the SQL with the number of DWs, $M$, it exceeds the classical bound as a function of $n$ [compare with Eq.~\eqref{eq.snl}]. The inter-DW mixing is nonlocal in terms
of DWs, hence it disturbes the structure of Eqs~\eqref{eq.css.mix} and~\eqref{eq.css.sep}. This improves the sensitivity through a many-body equivalent of the Hong-Ou-Mandel effect~\cite{hong,knight_qo}, which is
known in quantum optics. In our case, the mixing distributes the bosons across the neighboring DWs. The subsequent first beam-splitter opening the MZI plays the role analogue to the beam-splitter in the
two-photon HOM effect---interference of bosons yields a non-classical state~\cite{lopes2015atomic,identical_plenio,grangier,yurke,PhysRevA.91.043619}.

\begin{figure}[t!]
    \centering
    \includegraphics[width=1.0\linewidth]{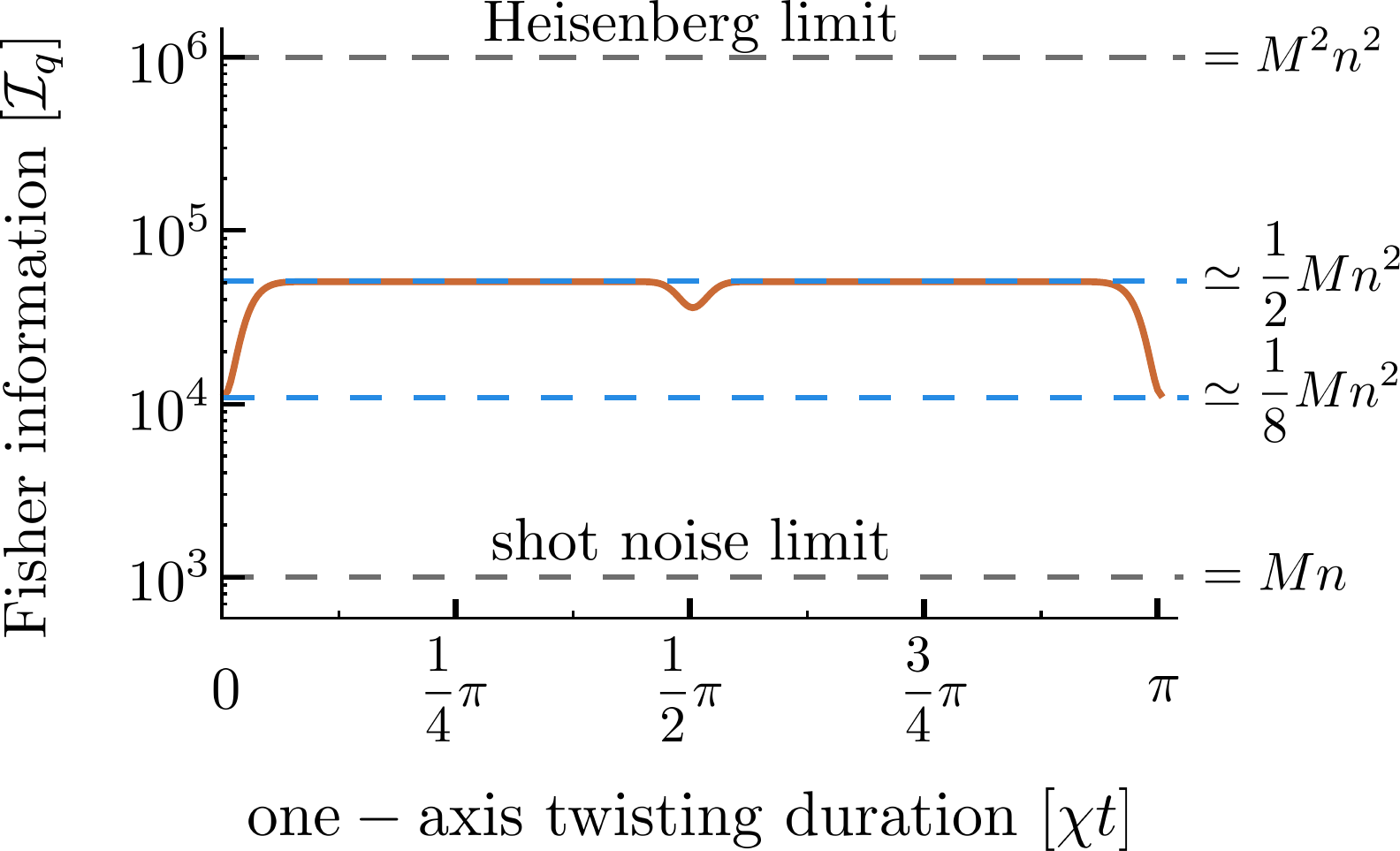}
    \caption{The QFI $\Iq$ calculated with $M=10$ DWs, each initially loaded with $n=100$ atoms. The gray dashed lines at the top and bottom show the HL and SQL, respectively.
      The two blue dashed lines in the center show the sensitivities that can be achieved using coherent states, as defined by Eq.~\eqref{eq.css.2} (upper) and Eq.~\eqref{eq.cssb} (lower line).
      The solid red line shows the QFI obtained using the output of the OAT procedure, Eq.~\eqref{eq.sss}, as a function of its duration $\chi t$.}
    \label{fig.pure}
\end{figure}

Another way of loading $M$ BECs into the array of DWs is by distributing each set of $n$ bosons coherently and symmetrically among the two sites. This corresponds to 
a product of single-DW CSSs, as in Eq.~\eqref{eq.css} with $\theta_i=\frac\pi2$ and $\varphi_i=0$ for all $i$, namely
\begin{align}\label{eq.cssb.one}
  \ket{\phi^{(i)}}&=\frac1{\sqrt{n!}}\left(\frac{\hat a_r^{(i)^\dagger}+\hat a_l^{(i)^\dagger}}{\sqrt2}\right)^n\ket0_i\nonumber\\
  &=\sum_{m_i=0}^n\sqrt{\frac{1}{2^n}\binom{n}{m_i}}\ket{m_i,n-m_i}_i,
\end{align}
giving the composite state
\begin{align}\label{eq.cssb}
  \ket\phi=\bigotimes^M_{i=1}\ket{\phi^{(i)}}.
\end{align}
Using Eq.~\eqref{eq.qfim} we obtain the QCRB [again, see Appendix~\ref{app1} for the details of the derivation]
\begin{align}
  \Iq=\frac18Mn^2+\frac12(1-\sqrt2)n^2+\frac14\left(\frac72M+1\right)n.
\end{align}
Also in this case, the sensitivity surpasses the SQL with the dominant (for large $n$ and $M$) scaling $\Iq\sim\frac18Mn^2$. 

\subsection{One-axis twisting}\label{sec.oat}

These two CSSs show that even when the input of the multi-DW interferometer is a set of independent semi-classical atomic states, the SQL can still be surpassed.  
For reference, we calculate the QCRB for the case when the input is composed of $M$ independent but entangled states. To control the strength of entanglement, we
consider the one-axis twisting protocol, where each state, initially as in Eq.~\eqref{eq.cssb.one}, undergoes the following dynamics
\begin{equation}\label{eq.sss.one}
  \ket{\psi^{(i)}(t)}=\sum^{n}_{m_i=0}e^{-i\chi t m_i^2}\sqrt{\frac{1}{2^n}\binom{n}{m_i}}\ket{m_i,n-m_i}_{i}.
\end{equation}
This type of evolution is generated by the two-body collisions quantified by the coupling constant $\chi$. Once the locally entangled state is prepared, the input of the mixing/MZI dynamics is
\begin{equation}\label{eq.sss}
  \ket{\psi(t)}=\bigotimes_{i=1}^M\ket{\psi^{(i)}(t)}.
\end{equation}
The QCRB calculated with this state and Eq.~\eqref{eq.qfim} gives a complicated formula which we reproduce in full extent in the Appendix~\ref{app1}. Nevertheless, at most of the times $t$ such that $\chi t$
is not an integer multiple of $\pi/2$, the $\Iq$ vastly simplifies as the quickly oscillating phase terms $e^{-i\chi t m_i^2}$ average to zero giving
\begin{equation}
  \Iq\sim\frac 12Mn^2.
\end{equation}
Hence in our scenario this complex entangling procedure does not give any improvement when compared to the performance of $M$ separable states, see Eq.~\eqref{eq.scale.fock}.
This observation could simplify future implementations of the multi-DW scheme discussed in this work.

To summarize the results of this section, in Fig.~\ref{fig.pure} we display the performance of the OAT state, as a function of $\chi t$ for $n=100$ and $M=10$. This is compared with the SQL and the
HL, see Eqs~\eqref{eq.snl} and~\eqref{eq.hl} respectively. The two other distinct values, $\Iq=\frac12Mn^2$ and $\Iq=\frac18Mn^2$, achievable with the CSSs are shown for reference. The take-away
message is that the use of CSS, accompanied by the inter-DW mixing, can yield very high sensitivity, thanks to the many-body HOM effect.

\subsection{Atom Number Fluctuations}\label{sec.fluct}

\begin{figure}[t!]
    \centering
    \includegraphics[width=1.0\linewidth]{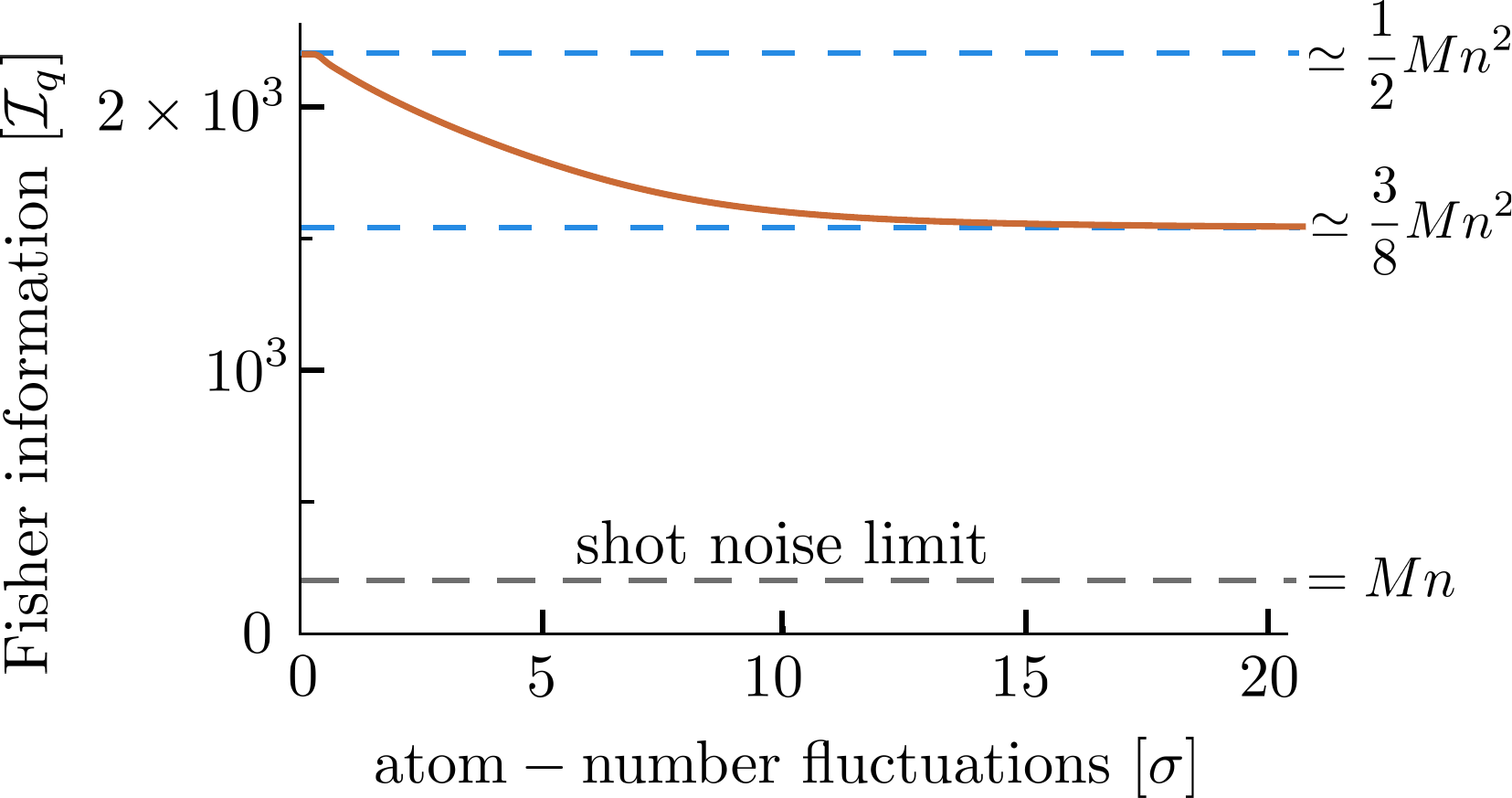}
    \caption{Red solid line; the QFI as a function of the strength of the shot-to-shot atom fluctuations $\sigma$, as given by Eq.~\eqref{eq.prob.sigma} with $n=20$ and $M=10$. 
      The dashed gray line at the bottom
      shows the SQL. The top dashed line denotes the value of $\Iq$ which is attained in the absence of fluctuations, see Eq.~\eqref{eq.scale.fock}. The central dashed line is the analytical result 
      for $\sigma\rightarrow\infty$, see Eq.~\eqref{eq.limit}.}
    \label{fig.mixed}
\end{figure}

Since the BEC loaded into just one site of a DW, Eq.~\eqref{eq.css.2}, gives a very good scaling with $n$, see Eq.~\eqref{eq.scale.fock}, we focus on this scenario and scrutinize the impact
of shot-to-shot atom fluctuations. Now, each DW is loaded with a mixture
\begin{align}\label{eq.dens.mat}
  \hat\varrho_i=\sum^{n}_{m_i=0}p(m_i)\ketbra{m_i,n-m_i}{m_i,n-m_i},
\end{align}
and, for illustration, we use the Gaussian probability
\begin{align}\label{eq.prob.sigma}
  p(m_i)=\mathcal Ne^{-\frac{m_i^2}{2\sigma^2}},
\end{align}
where $\mathcal N$ is the normalization constant that depends on the strength of the shot-to-shot fluctuations $\sigma$. Now, the BEC can be incoherently distributed among the two sites. 
We feed the mixing/MZI protocol with a product of $M=10$ such states, each with $n=20$ atoms and calculate the $\Iq$ numerically, using the general expression from Eq.~\eqref{eq.qfi.gen}. The
result is displayed in Fig.~\ref{fig.mixed} together with the SQL and the two limiting cases. One is for $\sigma\rightarrow0$, when the numerical computation 
recovers the formula from Eq.~\eqref{eq.scale.fock}. The other is for $\sigma\rightarrow\infty$, where the analytical result shows that
\begin{align}\label{eq.limit}
  \lim_{\sigma\rightarrow\infty}\Iq=\frac{n^2+2n}{8}(3M-2)\simeq\frac38Mn^2.
\end{align}
Hence even large shot-to-shot fluctuations of the BEC distribution among the two sites do not kill the sub shot-noise scaling of the QCRB. All the intermediate steps of the calculation leading to the above
result are presented in Appendix~\ref{app2}.

\subsection{Optimal measurement}\label{sec.measure}

In this final section we show that the measurement of the number of particles in each site and the subsequent estimation of $\theta$ from the probability distribution of these outcomes is optimal for
any input state that has real coefficients and when working around $\theta=0$. 
This applies to both the pure CSSs considered in the previous section.
By ``optimal'', we mean that the sensitivity for this particular estimation protocol saturates the QCRB. 

Let $\ket{\psi_{in}}$ be an input state of the MZI interferometer. We consider a general state and hence do not specify if the inter-DW mixing was performed. The probability of finding the output state
\begin{align}
  \ket{\psi(\theta)}=e^{-i\theta\hat J_y}\ket{\psi_{in}}
\end{align}
in any of the states from Eq.~\eqref{eq.vecn} is
\begin{align}
  p(\vec n|\theta)=\modsq{\braket{\vec n}{\psi(\theta)}}.
\end{align}
From this probability, the parameter $\theta$ can be estimated with the maximal sensitivity as in Eq.~\eqref{eq.qcrb} but with $\Id{\hat\varrho(\theta)}$ replaced by
\begin{align}\label{eq.class}
  \Icl=\sum_{\vec n}\frac1{p(\vec n|\theta)}\left(\frac{\partial p(\vec n|\theta)}{\partial\theta}\right)^2.
\end{align}
Taking $\theta=0$ as a working point, around which the derivative is calculated, we obtain that, for all the states $\ket{\psi_{in}}$ with real coefficients of
the expansion into the basis of $\ket{\vec n}$, namely $\braket{\vec n}{\psi_{in}}$, the probability is
\begin{align}
  p(\vec n|0)=\braket{\vec n}{\psi_{in}}^2.
\end{align}
For this case, the derivative of the probability is
\begin{align}
  \frac{\partial p(\vec n|\theta)}{\partial\theta}\Big|_{\theta=0}=i\braket{\vec n}{\psi_{in}}\left(\bra{\vec n}\hat J_y\ket{\psi_{in}}-\bra{\psi_{in}}\hat J_y\ket{\vec n}\right).
\end{align}
Plugging this expression into Eq.~\eqref{eq.class} we obtain
\begin{align}
  \Icl=\sum_{\vec n}\left(2\modsq{\bra{\vec n}\hat J_y\ket{\psi_{in}}}-\bra{\vec n}\hat J_y\ket{\psi_{in}}^2-\bra{\psi_{in}}\hat J_y\ket{\vec n}^2\right)
\end{align}
Taking the first term separately, we note that the sum over $\vec n$ can be performed giving
\begin{align}
  \sum_{\vec n}\modsq{\bra{\vec n}\hat J_y\ket{\psi_{in}}}=\av{\hat J_y^2},
\end{align}
where the average is calculated in the input state $\ket{\psi_{in}}$. As for the latter two terms, we have
\begin{align}
  &\bra{\vec n}\hat J_y\ket{\psi_{in}}^2+\bra{\psi_{in}}\hat J_y\ket{\vec n}^2=\\
  &=\left(\bra{\vec n}\hat J_y\ket{\psi_{in}}+\bra{\psi_{in}}\hat J_y\ket{\vec n}\right)^2-2\modsq{\bra{\vec n}\hat J_y\ket{\psi_{in}}}\nonumber.
\end{align}
Hence the CFI is equal to
\begin{align}
  \Icl=4\av{\hat J_y^2}-4\sum_{\vec n}\left(\re{\bra{\vec n}\hat J_y\ket{\psi_{in}}}\right)^2
\end{align}
Real coefficients of the input state imply that $\re{\bra{\vec n}\hat J_y\ket{\psi_{in}}}=0$ [due to the presence of the imaginary unit in the definition of $\hat J_y$, see Eq.~\eqref{eq.def.jy}],
hence we obtain that
\begin{align}
  \Icl=4\av{\hat J_y^2},
\end{align}
which saturates Eq.~\eqref{eq.qfi.pure}. Hence, the measurement of the population imbalance is optimal as the $\Icl$ saturates the QCRB for $\ket\psi$'s that have real coefficients of
the expansion in the basis of eigen-states of $\hat J_z$.

\section{Conclusions}

In this manuscript, we presented an analysis of a composite interferometer operating on a collection of DWs. We calculated the ultimate scaling with respect to the number of particles
and the number of DWs that can be achieved with this setup. 
Next, we introduced the inter-DW mixing which enables high sensitivity even when the array is loaded with a collection of semiclassical states. 
The many-body equivalent of the HOM effect is potent and resilient in the face of atomic fluctuations. Additionally, we demonstrated that the estimation obtained by measuring 
the number of atoms in each DW is optimal, meaning it saturates the QCRB.

The improvement from using CSSs and mixing is expressed in terms of an HL scaling with $n$. To obtain sub-SQL sensitivities in terms of $M$ as well, 
long-range coherence must be established in the system. For this purpose, other than nearest-neighbor beam-splitter operations should be considered.

Of course, the discussion presented here is only the first step in a complete analysis of the interferometric performance of this setup.
A more realistic scenario is one in which the atom-number measurements are imprecise. Phase noise can cause fluctuations in the relative depth of the DWs. 
The number of particles in each DW and the total number of particles can also fluctuate from shot to shot. Other sources of decoherence can further 
reduce precision. Nevertheless, we hope that the results presented in this work will inspire further research on this novel, promising interferometric scheme.

\section*{Ackonwledgements}

This work was supported by the National Science Centre, Poland, within the QuantERA II Programme that has received funding from the European Union’s Horizon 2020 
research and innovation programme under Grant Agreement No 101017733, Project No. 2021/03/Y/ST2/00195.

\onecolumngrid

\appendix

\section{QFI for a Pure state}\label{app1}
 The generator of the interferometric transformation is
$\hat J_y=\sum_{i=1}^M\hat J^{(i)}_{y}.$ Since all $\hat J^{(i)}_{y}$'s commute, we can write the QFI as follows
\begin{align}\label{eq.fmi}
  \Iq=4\sum_{i=1}^M\Delta^2\hat S^{(i)}_{y}+8\sum_{i\neq j=1}^M\left(\av{\hat S^{(i)}_{y}\hat S^{(j)}_{y}}-\av{\hat S^{(i)}_{y}}\av{\hat S^{(j)}_{y}}\right)
\end{align} 
The first part of the QFI simplifies to
\begin{align}
  4\sum_{i=1}^M\Delta^2\hat S^{(i)}_{y}&=2\Delta^2\hat J^{(1)}_{y}+2\Delta^2\hat J^{(M)}_{y}+2\av{\hat L^{(2)2}}+2\av{\hat R^{(M)2}}\nonumber
  +\\&+\sum_{i=2}^{M-1}\Delta^2\hat J^{(i)}_{y}+\sum_{i=2}^{M-1}\bigg(\av{\hat Y_y^{(i)2}}+\av{\hat L^{(i)2}}+\av{\hat R^{(i)2}}\bigg).
\end{align}
The second term also simplifies, leaving the only nonzero correlations between the neighbouring operators, like $\av{\hat L^{(i)} \hat R^{(i)}}$. Therefore this second part of
the QFI reads
\begin{align}
 8\sum_{i\neq j=1}^M\left(\av{\hat S^{(i)}_{y}\hat S^{(j)}_{y}}-\av{\hat S^{(i)}_{y}}\av{\hat S^{(j)}_{y}}\right)=-\frac4{\sqrt2}\av{\hat L^{(2)}\hat R^{(2)}}
  -\frac4{\sqrt2}\av{\hat L^{(M)}\hat R^{(M)}}-2\sum_{i=3}^{M-1}\av{\hat L^{(i)}\hat R^{(i)}}.
\end{align}
QFI for the state $\ket{\psi}$ using Eq. ({\ref{eq.fmi}}),
\begin{subequations}
  \begin{align}
    &2\Delta^2\hat J^{(1)}_{y}=2\Delta^2\hat J^{(M)}_{y}=\frac{n}{2},\ \ \ 2\av{\hat R^{(i)2}}=n(n+1),\ \ \ 2\av{\hat L^{(i)2}}=0\\
    &\sum^{M-1}_{i=2} \Delta^2\hat J^{(i)}_{y}=\sum^{M-1}_{i=2} \av{Y_y^{(i)}}=(M-2)\frac{n}{4}
  \end{align}
\end{subequations}
Hence the QFI becomes, $\Iq=\frac{M}{2}(n^2+2n)=\frac{N^2}{2M}+N$.

Using Eq. (\ref{eq.fmi}), QFI for the state becomes
\begin{subequations}
  \begin{align}
    & \Delta^2\hat J^{(i)}_{y}=\frac n4,\ \ \ \av{\hat Y^{(i)2}_{y}}=\frac n4(n+1),\ \ \ \av{\hat R^{(i)2}}=\av{\hat L^{(i)2}}=\frac n4 \big(\frac n2 +1 \big)\\
    &\av{\hat L^{(2)}\hat R^{(2)}}=\av{\hat L^{(i)}\hat R^{(i)}}=\av{\hat L^{(M)}\hat R^{(M)}}=\frac{n^2}8.
\end{align}
\end{subequations}
Hence the full expression for the QFI is
\begin{align}
    F_{q}=Mn(\frac{7+n}{8})-\frac{n^2}{\sqrt{2}}+\frac{n}{2}(n+\frac{1}{2})
\end{align}

QFI for the state $\ket{\psi_\alpha}$ using Eq. (\ref{eq.fmi}) has a general form
\begin{align}\label{eq.fqs}
  \Iq=n^2+\frac{3n}{2}+(M-2)\frac n2\left(n+\frac32\right)-\frac{M+2}{2}\re{f}-\left(M-3+\frac{4}{\sqrt{2}}\right)\modsq{g}
\end{align}
where the functions $f$ and $g$ are defined as follows
\begin{align}
  g&=\av{\hat{a}_{r}^{\dagger}\hat{a}_l}=\frac{1}{2^n}\sum^{n/2}_{m=-n/2}e^{-i\alpha (2m+1))}\binom{n}{\frac n2-m}\bigg(\frac n2-m\bigg)\\
  f&=\av{\hat{a}_r^{\dagger}\hat{a}_l\hat{a}_r^{\dagger}\hat{a}_l}=\frac{1}{2^n}\sum^{\frac n2}_{m=-\frac n2}e^{-i\alpha (2m^2+4m+4)}\times\nonumber\\
  &\times\sqrt{\binom{n}{\frac n2+m}\binom{n}{\frac n2+m+2}(\frac n2+m+1)\left(\frac n2-m\right)\left(\frac n2+m+2\right)\left(\frac n2-m-1\right)}
\end{align}

\section{QFI for a mixed state}\label{app2}
The density matrix in Eq.~\eqref{eq.dens.mat} has some zero eigen values, 
hence the QFI must be handled with care. By labelling the non-zero eigenvalues of the density matrix with $n$ and the zeros with $n'$, i.e.,
\begin{align}
  \hat\varrho=\sum_{\tilde n}p_{\tilde n}\ketbra{\tilde n}{\tilde n}=\sum_{n}p_{n}\ketbra nn+\sum_{n'}0\ketbra{n'}{n'}.
\end{align}
we obtain the expression for the QFI as follows
\begin{align}
  \mathcal{I}_q=2\sum_{\tilde n,\tilde m}\frac{(p_{\tilde n}-p_{\tilde m})^2}{p_{\tilde n}+p_{\tilde m}}\modsq{\bra{\tilde n}\hat J_y\ket{\tilde m}}=
  2\sum_{n,m}\frac{(p_{n}-p_{m})^2}{p_{n}+p_{m}}\modsq{\bra{n}\hat J_y\ket{m}}+4\sum_{n,m'}p_n\modsq{\bra{n}\hat J_y\ket{m'}}.
\end{align}
The first part, we denote it by $\mathcal{I}_q^{(1)}$ is the QFI for the non-zero subspace only. Consider the second part
\begin{align}
  &4\sum_{n,m'}p_n\modsq{\bra{n}\hat J_y\ket{m'}}=4\sum_{n}p_n\bra{n}\hat J_y\sum_{m'}\ketbra{m'}{m'}\hat J_y\ket n=4\sum_{n}p_n\bra{n}\hat J_y(\hat{\mathds{1}}-\sum_{m}\ketbra{m}{m})\hat J_y\ket n\nonumber\\
  &=4\sum_{n}p_n\bra{n}\hat J_y^2\ket n-4\sum_{n,m}p_n\modsq{\bra{n}\hat J_y\ket{m}}=4\tr{\hat\varrho\hat J_y^2}-4\sum_{n,m}p_n\modsq{\bra{n}\hat J_y\ket{m}}.
\end{align}
Hence the QFI is equal to
\begin{align}\label{eq.qfims}
  \mathcal{I}_q=\mathcal{I}_q^{(1)}+4\tr{\hat\varrho\hat J_y^2}-4\sum_{n,m}p_n\modsq{\bra{n}\hat J_y\ket{m}}.
\end{align}
we now calculate the three contributions separately
\begin{align}
     \mathcal{I}^{(1)}_{q}&=\sum_{ij}2\frac{(p_i-p_j)^2}{p_i+p_j}|\bra{\psi_i}\hat{S_y}\ket{\psi_j}|^2=\frac{1}{4}\bigg[\sum_{n_1,n'_1=0}\frac{(p_{n_1}-p_{n'_1})^2}{p_{n_1}+p_{n'_1}}\modsq{d[n_1,n'_1]}\nonumber\\
     &+\sum_{n_M,n'_M=0}\frac{(p_{n_M}-p_{n'_M})^2}{p_{n_M}+p_{n'_M}}\modsq{d[n_M,n'_M]}+\sum_{n_i,n'_i=0}\frac{(p_{n_i}-p_{n'_i})^2}{p_{n_i}-p_{n'_i}}\frac{(M-2)}{2}\modsq{d[n_i,n'_i]}\bigg].
\end{align}
For clarity we use the notation $\bra{n'_i,n-n'_i}\hat S_{y}\ket{n_i,n-n_i}\equiv\bra{\psi'_i}\hat S_y\ket{\psi_i}$ where,
\begin{align}
 \bra{\psi'_i}\hat S_y\ket{\psi_i}=\frac{1}{2i}\big(\sqrt{(n-n_i)(n_i+1)}\delta_{n'_i=n_i+1}- \sqrt{(n-n_i+1)n_i}\delta_{n'_i=n_i-1}\big)=\frac{1}{2i}d[n_i,n'_i].
\end{align}
The second term is
\begin{align}
  \mathcal{I}^{(2)}_{q}&=\sum_{i}4p_{n_i}\bra{\psi_i}\hat{S_y}^2\ket{\psi_i}=\frac{1}{2}\sum^{n}_{n_1,n_2,...n_M}p_{n_1}p_{n_2},...p_{n_M}\times\\
  &\times\bigg[\bigg(2n(n+2)+2nn_1-2n^{2}_{M}-2n^{2}_{1}+2n_1n_2+n_1+n_2-2nn_{M-1}+2n_{M}n_{M-1}-n_{M-1}-n_M\bigg)\nonumber\\
    &+\frac{(M-2)}{2}\bigg(2n(n+2)-2n^{2}_{i}+2n_i n_{i+1}+2n_{i+1}-2n_{i-1}-2nn_{i-1}+2nn_{i+1}+2n_{i-1}n_{i}-2n_{i-1}n_{i+1} \bigg)   \bigg]\nonumber 
\end{align}
The last term reads
\begin{align}\label{maf.fq.3}
  \mathcal{I}^{(3)}_{q}&=-\sum_{ij}4p_{n_i}|\bra{\psi_i}\hat{S_y}\ket{\psi_j}|^2=-\frac 12\bigg[\sum_{n_1,n'_1=0}p_{n_1}\modsq{d[n_1,n'_1]}\nonumber\\
    &+\sum_{n_M,n'_M=0}p_{n_M}\modsq{d[n_M,n'_M]}+\sum_{n_k,n'_k=0}p_{n_k}\frac{(M-2)}{2}\modsq{d[n_i,n'_i]}\bigg]
\end{align}
Furthermore, we observe that the first term $\mathcal{I}^{(1)}_q$ vanishes. The terms $\mathcal{I}_{q}^{(2)}$ and $\mathcal{I}_{q}^{(3)}$ results for large $\sigma$ in 
\begin{align}\label{eq.qfifp}
  \Iq&=\bigg(\frac{11M-2}{12}\bigg)\bigg(\frac{n^2}{2}+n\bigg)-\frac{1}{2(n+1)}\bigg(\sum_{n_1,n'_1=0}|d[n_1,n'_1]|^2\nonumber\\
  &+\sum_{n_M,n'_M=0}|d[n_M,n'_M]|^2+\frac{M-2}{2}\sum_{n_i,n'_i=0}|d[n_i,n'_i]|^2\bigg).
\end{align}
The summations can be performed analytically giving
\begin{align}
  \sum_{n_1,n'_1=0}\big|d[n_1,n'_1]\big|^2=\sum_{n_M,n'_M=0}\big|d[n_M,n'_M]\big|^2=\sum_{n_i,n'_i=0}\big|d[n_i,n'_i]\big|^2=\frac{1}{3}(n^3+3n^2+2n),
\end{align}
resulting in a simplified expression for the QFI,
\begin{align}
  \mathcal{I}_q(\sigma\rightarrow\infty)=\frac{n^2+2n}{8}(3M-2),
\end{align}
as reported in the main text.

\twocolumngrid


\begin{thebibliography}{55}%
\makeatletter
\providecommand \@ifxundefined [1]{%
 \@ifx{#1\undefined}
}%
\providecommand \@ifnum [1]{%
 \ifnum #1\expandafter \@firstoftwo
 \else \expandafter \@secondoftwo
 \fi
}%
\providecommand \@ifx [1]{%
 \ifx #1\expandafter \@firstoftwo
 \else \expandafter \@secondoftwo
 \fi
}%
\providecommand \natexlab [1]{#1}%
\providecommand \enquote  [1]{``#1''}%
\providecommand \bibnamefont  [1]{#1}%
\providecommand \bibfnamefont [1]{#1}%
\providecommand \citenamefont [1]{#1}%
\providecommand \href@noop [0]{\@secondoftwo}%
\providecommand \href [0]{\begingroup \@sanitize@url \@href}%
\providecommand \@href[1]{\@@startlink{#1}\@@href}%
\providecommand \@@href[1]{\endgroup#1\@@endlink}%
\providecommand \@sanitize@url [0]{\catcode `\\12\catcode `\$12\catcode
  `\&12\catcode `\#12\catcode `\^12\catcode `\_12\catcode `\%12\relax}%
\providecommand \@@startlink[1]{}%
\providecommand \@@endlink[0]{}%
\providecommand \url  [0]{\begingroup\@sanitize@url \@url }%
\providecommand \@url [1]{\endgroup\@href {#1}{\urlprefix }}%
\providecommand \urlprefix  [0]{URL }%
\providecommand \Eprint [0]{\href }%
\providecommand \doibase [0]{https://doi.org/}%
\providecommand \selectlanguage [0]{\@gobble}%
\providecommand \bibinfo  [0]{\@secondoftwo}%
\providecommand \bibfield  [0]{\@secondoftwo}%
\providecommand \translation [1]{[#1]}%
\providecommand \BibitemOpen [0]{}%
\providecommand \bibitemStop [0]{}%
\providecommand \bibitemNoStop [0]{.\EOS\space}%
\providecommand \EOS [0]{\spacefactor3000\relax}%
\providecommand \BibitemShut  [1]{\csname bibitem#1\endcsname}%
\let\auto@bib@innerbib\@empty
\bibitem [{\citenamefont {Braunstein}\ and\ \citenamefont
  {Caves}(1994)}]{braunstein1994statistical}%
  \BibitemOpen
  \bibfield  {author} {\bibinfo {author} {\bibfnamefont {S.~L.}\ \bibnamefont
  {Braunstein}}\ and\ \bibinfo {author} {\bibfnamefont {C.~M.}\ \bibnamefont
  {Caves}},\ }\bibfield  {title} {\bibinfo {title} {Statistical distance and
  the geometry of quantum states},\ }\href@noop {} {\bibfield  {journal}
  {\bibinfo  {journal} {Physical Review Letters}\ }\textbf {\bibinfo {volume}
  {72}},\ \bibinfo {pages} {3439} (\bibinfo {year} {1994})}\BibitemShut
  {NoStop}%
\bibitem [{\citenamefont {Holevo}(2011)}]{holevo2011probabilistic}%
  \BibitemOpen
  \bibfield  {author} {\bibinfo {author} {\bibfnamefont {A.}~\bibnamefont
  {Holevo}},\ }\href@noop {} {\emph {\bibinfo {title} {Probabilistic and
  Statistical Aspects of Quantum Theory}}}\ (\bibinfo  {publisher}
  {Publications of Scuola Normale Superiore},\ \bibinfo {year}
  {2011})\BibitemShut {NoStop}%
\bibitem [{\citenamefont {Giovannetti}\ \emph {et~al.}(2004)\citenamefont
  {Giovannetti}, \citenamefont {Lloyd},\ and\ \citenamefont
  {Maccone}}]{giovannetti2004quantum}%
  \BibitemOpen
  \bibfield  {author} {\bibinfo {author} {\bibfnamefont {V.}~\bibnamefont
  {Giovannetti}}, \bibinfo {author} {\bibfnamefont {S.}~\bibnamefont {Lloyd}},\
  and\ \bibinfo {author} {\bibfnamefont {L.}~\bibnamefont {Maccone}},\
  }\bibfield  {title} {\bibinfo {title} {Quantum-enhanced measurements: beating
  the standard quantum limit},\ }\href@noop {} {\bibfield  {journal} {\bibinfo
  {journal} {Science}\ }\textbf {\bibinfo {volume} {306}},\ \bibinfo {pages}
  {1330} (\bibinfo {year} {2004})}\BibitemShut {NoStop}%
\bibitem [{\citenamefont {Abbott}\ \emph {et~al.}(2016)\citenamefont {Abbott},
  \citenamefont {Abbott}, \citenamefont {Abbott}, \citenamefont {Abernathy},
  \citenamefont {Acernese}, \citenamefont {Ackley}, \citenamefont {Adams},
  \citenamefont {Adams}, \citenamefont {Addesso}, \citenamefont {Adhikari}
  \emph {et~al.}}]{abbott2016observation}%
  \BibitemOpen
  \bibfield  {author} {\bibinfo {author} {\bibfnamefont {B.~P.}\ \bibnamefont
  {Abbott}}, \bibinfo {author} {\bibfnamefont {R.}~\bibnamefont {Abbott}},
  \bibinfo {author} {\bibfnamefont {T.~D.}\ \bibnamefont {Abbott}}, \bibinfo
  {author} {\bibfnamefont {M.~R.}\ \bibnamefont {Abernathy}}, \bibinfo {author}
  {\bibfnamefont {F.}~\bibnamefont {Acernese}}, \bibinfo {author}
  {\bibfnamefont {K.}~\bibnamefont {Ackley}}, \bibinfo {author} {\bibfnamefont
  {C.}~\bibnamefont {Adams}}, \bibinfo {author} {\bibfnamefont
  {T.}~\bibnamefont {Adams}}, \bibinfo {author} {\bibfnamefont
  {P.}~\bibnamefont {Addesso}}, \bibinfo {author} {\bibfnamefont {R.~X.}\
  \bibnamefont {Adhikari}}, \emph {et~al.},\ }\bibfield  {title} {\bibinfo
  {title} {Observation of gravitational waves from a binary black hole
  merger},\ }\href@noop {} {\bibfield  {journal} {\bibinfo  {journal} {Physical
  review letters}\ }\textbf {\bibinfo {volume} {116}},\ \bibinfo {pages}
  {061102} (\bibinfo {year} {2016})}\BibitemShut {NoStop}%
\bibitem [{\citenamefont {Ferrari}\ \emph {et~al.}(2006)\citenamefont
  {Ferrari}, \citenamefont {Poli}, \citenamefont {Sorrentino},\ and\
  \citenamefont {Tino}}]{ferrari}%
  \BibitemOpen
  \bibfield  {author} {\bibinfo {author} {\bibfnamefont {G.}~\bibnamefont
  {Ferrari}}, \bibinfo {author} {\bibfnamefont {N.}~\bibnamefont {Poli}},
  \bibinfo {author} {\bibfnamefont {F.}~\bibnamefont {Sorrentino}},\ and\
  \bibinfo {author} {\bibfnamefont {G.~M.}\ \bibnamefont {Tino}},\ }\bibfield
  {title} {\bibinfo {title} {Long-lived bloch oscillations with bosonic sr
  atoms and application to gravity measurement at the micrometer scale},\
  }\href {https://doi.org/10.1103/PhysRevLett.97.060402} {\bibfield  {journal}
  {\bibinfo  {journal} {Phys. Rev. Lett.}\ }\textbf {\bibinfo {volume} {97}},\
  \bibinfo {pages} {060402} (\bibinfo {year} {2006})}\BibitemShut {NoStop}%
\bibitem [{\citenamefont {Poli}\ \emph {et~al.}(2011)\citenamefont {Poli},
  \citenamefont {Wang}, \citenamefont {Tarallo}, \citenamefont {Alberti},
  \citenamefont {Prevedelli},\ and\ \citenamefont {Tino}}]{poli}%
  \BibitemOpen
  \bibfield  {author} {\bibinfo {author} {\bibfnamefont {N.}~\bibnamefont
  {Poli}}, \bibinfo {author} {\bibfnamefont {F.-Y.}\ \bibnamefont {Wang}},
  \bibinfo {author} {\bibfnamefont {M.~G.}\ \bibnamefont {Tarallo}}, \bibinfo
  {author} {\bibfnamefont {A.}~\bibnamefont {Alberti}}, \bibinfo {author}
  {\bibfnamefont {M.}~\bibnamefont {Prevedelli}},\ and\ \bibinfo {author}
  {\bibfnamefont {G.~M.}\ \bibnamefont {Tino}},\ }\bibfield  {title} {\bibinfo
  {title} {Precision measurement of gravity with cold atoms in an optical
  lattice and comparison with a classical gravimeter},\ }\href
  {https://doi.org/10.1103/PhysRevLett.106.038501} {\bibfield  {journal}
  {\bibinfo  {journal} {Phys. Rev. Lett.}\ }\textbf {\bibinfo {volume} {106}},\
  \bibinfo {pages} {038501} (\bibinfo {year} {2011})}\BibitemShut {NoStop}%
\bibitem [{\citenamefont {Biedermann}\ \emph {et~al.}(2015)\citenamefont
  {Biedermann}, \citenamefont {Wu}, \citenamefont {Deslauriers}, \citenamefont
  {Roy}, \citenamefont {Mahadeswaraswamy},\ and\ \citenamefont
  {Kasevich}}]{PhysRevA.91.033629}%
  \BibitemOpen
  \bibfield  {author} {\bibinfo {author} {\bibfnamefont {G.~W.}\ \bibnamefont
  {Biedermann}}, \bibinfo {author} {\bibfnamefont {X.}~\bibnamefont {Wu}},
  \bibinfo {author} {\bibfnamefont {L.}~\bibnamefont {Deslauriers}}, \bibinfo
  {author} {\bibfnamefont {S.}~\bibnamefont {Roy}}, \bibinfo {author}
  {\bibfnamefont {C.}~\bibnamefont {Mahadeswaraswamy}},\ and\ \bibinfo {author}
  {\bibfnamefont {M.~A.}\ \bibnamefont {Kasevich}},\ }\bibfield  {title}
  {\bibinfo {title} {Testing gravity with cold-atom interferometers},\ }\href
  {https://doi.org/10.1103/PhysRevA.91.033629} {\bibfield  {journal} {\bibinfo
  {journal} {Phys. Rev. A}\ }\textbf {\bibinfo {volume} {91}},\ \bibinfo
  {pages} {033629} (\bibinfo {year} {2015})}\BibitemShut {NoStop}%
\bibitem [{\citenamefont {Dubetsky}\ and\ \citenamefont
  {Kasevich}(2006)}]{PhysRevA.74.023615}%
  \BibitemOpen
  \bibfield  {author} {\bibinfo {author} {\bibfnamefont {B.}~\bibnamefont
  {Dubetsky}}\ and\ \bibinfo {author} {\bibfnamefont {M.~A.}\ \bibnamefont
  {Kasevich}},\ }\bibfield  {title} {\bibinfo {title} {Atom interferometer as a
  selective sensor of rotation or gravity},\ }\href
  {https://doi.org/10.1103/PhysRevA.74.023615} {\bibfield  {journal} {\bibinfo
  {journal} {Phys. Rev. A}\ }\textbf {\bibinfo {volume} {74}},\ \bibinfo
  {pages} {023615} (\bibinfo {year} {2006})}\BibitemShut {NoStop}%
\bibitem [{\citenamefont {McGuirk}\ \emph {et~al.}(2002)\citenamefont
  {McGuirk}, \citenamefont {Foster}, \citenamefont {Fixler}, \citenamefont
  {Snadden},\ and\ \citenamefont {Kasevich}}]{PhysRevA.65.033608}%
  \BibitemOpen
  \bibfield  {author} {\bibinfo {author} {\bibfnamefont {J.~M.}\ \bibnamefont
  {McGuirk}}, \bibinfo {author} {\bibfnamefont {G.~T.}\ \bibnamefont {Foster}},
  \bibinfo {author} {\bibfnamefont {J.~B.}\ \bibnamefont {Fixler}}, \bibinfo
  {author} {\bibfnamefont {M.~J.}\ \bibnamefont {Snadden}},\ and\ \bibinfo
  {author} {\bibfnamefont {M.~A.}\ \bibnamefont {Kasevich}},\ }\bibfield
  {title} {\bibinfo {title} {Sensitive absolute-gravity gradiometry using atom
  interferometry},\ }\href {https://doi.org/10.1103/PhysRevA.65.033608}
  {\bibfield  {journal} {\bibinfo  {journal} {Phys. Rev. A}\ }\textbf {\bibinfo
  {volume} {65}},\ \bibinfo {pages} {033608} (\bibinfo {year}
  {2002})}\BibitemShut {NoStop}%
\bibitem [{\citenamefont {Kasevich}\ and\ \citenamefont
  {Chu}(1992)}]{kasevich1992measurement}%
  \BibitemOpen
  \bibfield  {author} {\bibinfo {author} {\bibfnamefont {M.}~\bibnamefont
  {Kasevich}}\ and\ \bibinfo {author} {\bibfnamefont {S.}~\bibnamefont {Chu}},\
  }\bibfield  {title} {\bibinfo {title} {Measurement of the gravitational
  acceleration of an atom with a light-pulse atom interferometer},\ }\href@noop
  {} {\bibfield  {journal} {\bibinfo  {journal} {Applied Physics B}\ }\textbf
  {\bibinfo {volume} {54}},\ \bibinfo {pages} {321} (\bibinfo {year}
  {1992})}\BibitemShut {NoStop}%
\bibitem [{\citenamefont {Fixler}\ \emph {et~al.}(2007)\citenamefont {Fixler},
  \citenamefont {Foster}, \citenamefont {McGuirk},\ and\ \citenamefont
  {Kasevich}}]{doi:10.1126/science.1135459}%
  \BibitemOpen
  \bibfield  {author} {\bibinfo {author} {\bibfnamefont {J.~B.}\ \bibnamefont
  {Fixler}}, \bibinfo {author} {\bibfnamefont {G.~T.}\ \bibnamefont {Foster}},
  \bibinfo {author} {\bibfnamefont {J.~M.}\ \bibnamefont {McGuirk}},\ and\
  \bibinfo {author} {\bibfnamefont {M.~A.}\ \bibnamefont {Kasevich}},\
  }\bibfield  {title} {\bibinfo {title} {Atom interferometer measurement of the
  newtonian constant of gravity},\ }\href
  {https://doi.org/10.1126/science.1135459} {\bibfield  {journal} {\bibinfo
  {journal} {Science}\ }\textbf {\bibinfo {volume} {315}},\ \bibinfo {pages}
  {74} (\bibinfo {year} {2007})},\ \Eprint
  {https://arxiv.org/abs/https://www.science.org/doi/pdf/10.1126/science.1135459}
  {https://www.science.org/doi/pdf/10.1126/science.1135459} \BibitemShut
  {NoStop}%
\bibitem [{\citenamefont {Fattori}\ \emph {et~al.}(2008)\citenamefont
  {Fattori}, \citenamefont {D'Errico}, \citenamefont {Roati}, \citenamefont
  {Zaccanti}, \citenamefont {Jona-Lasinio}, \citenamefont {Modugno},
  \citenamefont {Inguscio},\ and\ \citenamefont {Modugno}}]{fattori}%
  \BibitemOpen
  \bibfield  {author} {\bibinfo {author} {\bibfnamefont {M.}~\bibnamefont
  {Fattori}}, \bibinfo {author} {\bibfnamefont {C.}~\bibnamefont {D'Errico}},
  \bibinfo {author} {\bibfnamefont {G.}~\bibnamefont {Roati}}, \bibinfo
  {author} {\bibfnamefont {M.}~\bibnamefont {Zaccanti}}, \bibinfo {author}
  {\bibfnamefont {M.}~\bibnamefont {Jona-Lasinio}}, \bibinfo {author}
  {\bibfnamefont {M.}~\bibnamefont {Modugno}}, \bibinfo {author} {\bibfnamefont
  {M.}~\bibnamefont {Inguscio}},\ and\ \bibinfo {author} {\bibfnamefont
  {G.}~\bibnamefont {Modugno}},\ }\bibfield  {title} {\bibinfo {title} {Atom
  interferometry with a weakly interacting bose-einstein condensate},\
  }\href@noop {} {\bibfield  {journal} {\bibinfo  {journal} {Phys. Rev. Lett.}\
  }\textbf {\bibinfo {volume} {100}},\ \bibinfo {pages} {080405} (\bibinfo
  {year} {2008})}\BibitemShut {NoStop}%
\bibitem [{\citenamefont {Jachymski}\ \emph {et~al.}(2018)\citenamefont
  {Jachymski}, \citenamefont {Wasak}, \citenamefont {Idziaszek}, \citenamefont
  {Julienne}, \citenamefont {Negretti},\ and\ \citenamefont
  {Calarco}}]{PhysRevLett.120.013401}%
  \BibitemOpen
  \bibfield  {author} {\bibinfo {author} {\bibfnamefont {K.}~\bibnamefont
  {Jachymski}}, \bibinfo {author} {\bibfnamefont {T.}~\bibnamefont {Wasak}},
  \bibinfo {author} {\bibfnamefont {Z.}~\bibnamefont {Idziaszek}}, \bibinfo
  {author} {\bibfnamefont {P.~S.}\ \bibnamefont {Julienne}}, \bibinfo {author}
  {\bibfnamefont {A.}~\bibnamefont {Negretti}},\ and\ \bibinfo {author}
  {\bibfnamefont {T.}~\bibnamefont {Calarco}},\ }\bibfield  {title} {\bibinfo
  {title} {Single-atom transistor as a precise magnetic field sensor},\ }\href
  {https://doi.org/10.1103/PhysRevLett.120.013401} {\bibfield  {journal}
  {\bibinfo  {journal} {Phys. Rev. Lett.}\ }\textbf {\bibinfo {volume} {120}},\
  \bibinfo {pages} {013401} (\bibinfo {year} {2018})}\BibitemShut {NoStop}%
\bibitem [{\citenamefont {Dunningham}\ and\ \citenamefont
  {Burnett}(2004)}]{dunn2}%
  \BibitemOpen
  \bibfield  {author} {\bibinfo {author} {\bibfnamefont {J.~A.}\ \bibnamefont
  {Dunningham}}\ and\ \bibinfo {author} {\bibfnamefont {K.}~\bibnamefont
  {Burnett}},\ }\bibfield  {title} {\bibinfo {title} {Sub-shot-noise-limited
  measurements with bose-einstein condensates},\ }\href@noop {} {\bibfield
  {journal} {\bibinfo  {journal} {Phys. Rev. A}\ }\textbf {\bibinfo {volume}
  {70}},\ \bibinfo {pages} {033601} (\bibinfo {year} {2004})}\BibitemShut
  {NoStop}%
\bibitem [{\citenamefont {Pezz{\'e}}\ and\ \citenamefont
  {Smerzi}(2009)}]{pezze2009entanglement}%
  \BibitemOpen
  \bibfield  {author} {\bibinfo {author} {\bibfnamefont {L.}~\bibnamefont
  {Pezz{\'e}}}\ and\ \bibinfo {author} {\bibfnamefont {A.}~\bibnamefont
  {Smerzi}},\ }\bibfield  {title} {\bibinfo {title} {{Entanglement, nonlinear
  dynamics, and the Heisenberg limit}},\ }\href@noop {} {\bibfield  {journal}
  {\bibinfo  {journal} {Phys. Rev. Lett.}\ }\textbf {\bibinfo {volume} {102}},\
  \bibinfo {pages} {100401} (\bibinfo {year} {2009})}\BibitemShut {NoStop}%
\bibitem [{\citenamefont {Yadin}\ \emph {et~al.}(2021)\citenamefont {Yadin},
  \citenamefont {Fadel},\ and\ \citenamefont
  {Gessner}}]{yadin2021metrological}%
  \BibitemOpen
  \bibfield  {author} {\bibinfo {author} {\bibfnamefont {B.}~\bibnamefont
  {Yadin}}, \bibinfo {author} {\bibfnamefont {M.}~\bibnamefont {Fadel}},\ and\
  \bibinfo {author} {\bibfnamefont {M.}~\bibnamefont {Gessner}},\ }\bibfield
  {title} {\bibinfo {title} {Metrological complementarity reveals the
  einstein-podolsky-rosen paradox},\ }\href@noop {} {\bibfield  {journal}
  {\bibinfo  {journal} {Nature communications}\ }\textbf {\bibinfo {volume}
  {12}},\ \bibinfo {pages} {2410} (\bibinfo {year} {2021})}\BibitemShut
  {NoStop}%
\bibitem [{\citenamefont {Niezgoda}\ and\ \citenamefont
  {Chwede\ifmmode~\acute{n}\else
  \'{n}\fi{}czuk}(2021)}]{PhysRevLett.126.210506}%
  \BibitemOpen
  \bibfield  {author} {\bibinfo {author} {\bibfnamefont {A.}~\bibnamefont
  {Niezgoda}}\ and\ \bibinfo {author} {\bibfnamefont {J.}~\bibnamefont
  {Chwede\ifmmode~\acute{n}\else \'{n}\fi{}czuk}},\ }\bibfield  {title}
  {\bibinfo {title} {Many-body nonlocality as a resource for quantum-enhanced
  metrology},\ }\href {https://doi.org/10.1103/PhysRevLett.126.210506}
  {\bibfield  {journal} {\bibinfo  {journal} {Phys. Rev. Lett.}\ }\textbf
  {\bibinfo {volume} {126}},\ \bibinfo {pages} {210506} (\bibinfo {year}
  {2021})}\BibitemShut {NoStop}%
\bibitem [{\citenamefont {Fr\"owis}\ \emph {et~al.}(2019)\citenamefont
  {Fr\"owis}, \citenamefont {Fadel}, \citenamefont {Treutlein}, \citenamefont
  {Gisin},\ and\ \citenamefont {Brunner}}]{PhysRevA.99.040101}%
  \BibitemOpen
  \bibfield  {author} {\bibinfo {author} {\bibfnamefont {F.}~\bibnamefont
  {Fr\"owis}}, \bibinfo {author} {\bibfnamefont {M.}~\bibnamefont {Fadel}},
  \bibinfo {author} {\bibfnamefont {P.}~\bibnamefont {Treutlein}}, \bibinfo
  {author} {\bibfnamefont {N.}~\bibnamefont {Gisin}},\ and\ \bibinfo {author}
  {\bibfnamefont {N.}~\bibnamefont {Brunner}},\ }\bibfield  {title} {\bibinfo
  {title} {Does large quantum fisher information imply bell correlations?},\
  }\href {https://doi.org/10.1103/PhysRevA.99.040101} {\bibfield  {journal}
  {\bibinfo  {journal} {Phys. Rev. A}\ }\textbf {\bibinfo {volume} {99}},\
  \bibinfo {pages} {040101} (\bibinfo {year} {2019})}\BibitemShut {NoStop}%
\bibitem [{\citenamefont {Gerry}\ and\ \citenamefont
  {Campos}(2003)}]{PhysRevA.68.025602}%
  \BibitemOpen
  \bibfield  {author} {\bibinfo {author} {\bibfnamefont {C.~C.}\ \bibnamefont
  {Gerry}}\ and\ \bibinfo {author} {\bibfnamefont {R.~A.}\ \bibnamefont
  {Campos}},\ }\bibfield  {title} {\bibinfo {title} {Generation of maximally
  entangled states of a bose-einstein condensate and heisenberg-limited phase
  resolution},\ }\href {https://doi.org/10.1103/PhysRevA.68.025602} {\bibfield
  {journal} {\bibinfo  {journal} {Phys. Rev. A}\ }\textbf {\bibinfo {volume}
  {68}},\ \bibinfo {pages} {025602} (\bibinfo {year} {2003})}\BibitemShut
  {NoStop}%
\bibitem [{\citenamefont {Demkowicz-Dobrza\'nski}\ \emph
  {et~al.}(2012)\citenamefont {Demkowicz-Dobrza\'nski}, \citenamefont
  {Ko{\l}ody\'nski},\ and\ \citenamefont {Guta}}]{dobrz}%
  \BibitemOpen
  \bibfield  {author} {\bibinfo {author} {\bibfnamefont {R.}~\bibnamefont
  {Demkowicz-Dobrza\'nski}}, \bibinfo {author} {\bibfnamefont {J.}~\bibnamefont
  {Ko{\l}ody\'nski}},\ and\ \bibinfo {author} {\bibfnamefont {M.}~\bibnamefont
  {Guta}},\ }\bibfield  {title} {\bibinfo {title} {The elusive heisenberg limit
  in quantum-enhanced metrology},\ }\href@noop {} {\bibfield  {journal}
  {\bibinfo  {journal} {Nat. Commun.}\ }\textbf {\bibinfo {volume} {3}},\
  \bibinfo {pages} {1063} (\bibinfo {year} {2012})}\BibitemShut {NoStop}%
\bibitem [{\citenamefont {Schumm}\ \emph {et~al.}(2005)\citenamefont {Schumm},
  \citenamefont {Hofferberth}, \citenamefont {Andersson}, \citenamefont
  {Wildermuth}, \citenamefont {Groth}, \citenamefont {Bar-Joseph},
  \citenamefont {Schmiedmayer},\ and\ \citenamefont
  {Kr{\"u}ger}}]{schumm2005matter}%
  \BibitemOpen
  \bibfield  {author} {\bibinfo {author} {\bibfnamefont {T.}~\bibnamefont
  {Schumm}}, \bibinfo {author} {\bibfnamefont {S.}~\bibnamefont {Hofferberth}},
  \bibinfo {author} {\bibfnamefont {L.~M.}\ \bibnamefont {Andersson}}, \bibinfo
  {author} {\bibfnamefont {S.}~\bibnamefont {Wildermuth}}, \bibinfo {author}
  {\bibfnamefont {S.}~\bibnamefont {Groth}}, \bibinfo {author} {\bibfnamefont
  {I.}~\bibnamefont {Bar-Joseph}}, \bibinfo {author} {\bibfnamefont
  {J.}~\bibnamefont {Schmiedmayer}},\ and\ \bibinfo {author} {\bibfnamefont
  {P.}~\bibnamefont {Kr{\"u}ger}},\ }\bibfield  {title} {\bibinfo {title}
  {Matter-wave interferometry in a double well on an atom chip},\ }\href@noop
  {} {\bibfield  {journal} {\bibinfo  {journal} {Nat. Phys.}\ }\textbf
  {\bibinfo {volume} {1}},\ \bibinfo {pages} {57} (\bibinfo {year}
  {2005})}\BibitemShut {NoStop}%
\bibitem [{\citenamefont {Jo}\ \emph {et~al.}(2007)\citenamefont {Jo},
  \citenamefont {Shin}, \citenamefont {Will}, \citenamefont {Pasquini},
  \citenamefont {Saba}, \citenamefont {Ketterle}, \citenamefont {Pritchard},
  \citenamefont {Vengalattore},\ and\ \citenamefont {Prentiss}}]{jo2007long}%
  \BibitemOpen
  \bibfield  {author} {\bibinfo {author} {\bibfnamefont {G.-B.}\ \bibnamefont
  {Jo}}, \bibinfo {author} {\bibfnamefont {Y.}~\bibnamefont {Shin}}, \bibinfo
  {author} {\bibfnamefont {S.}~\bibnamefont {Will}}, \bibinfo {author}
  {\bibfnamefont {T.}~\bibnamefont {Pasquini}}, \bibinfo {author}
  {\bibfnamefont {M.}~\bibnamefont {Saba}}, \bibinfo {author} {\bibfnamefont
  {W.}~\bibnamefont {Ketterle}}, \bibinfo {author} {\bibfnamefont
  {D.}~\bibnamefont {Pritchard}}, \bibinfo {author} {\bibfnamefont
  {M.}~\bibnamefont {Vengalattore}},\ and\ \bibinfo {author} {\bibfnamefont
  {M.}~\bibnamefont {Prentiss}},\ }\bibfield  {title} {\bibinfo {title} {Long
  phase coherence time and number squeezing of two bose-einstein condensates on
  an atom chip},\ }\href@noop {} {\bibfield  {journal} {\bibinfo  {journal}
  {Phys. Rev. Lett.}\ }\textbf {\bibinfo {volume} {98}},\ \bibinfo {pages}
  {030407} (\bibinfo {year} {2007})}\BibitemShut {NoStop}%
\bibitem [{\citenamefont {B{\"o}hi}\ \emph {et~al.}(2009)\citenamefont
  {B{\"o}hi}, \citenamefont {Riedel}, \citenamefont {Hoffrogge}, \citenamefont
  {Reichel}, \citenamefont {H{\"a}nsch},\ and\ \citenamefont
  {Treutlein}}]{bohi2009coherent}%
  \BibitemOpen
  \bibfield  {author} {\bibinfo {author} {\bibfnamefont {P.}~\bibnamefont
  {B{\"o}hi}}, \bibinfo {author} {\bibfnamefont {M.~F.}\ \bibnamefont
  {Riedel}}, \bibinfo {author} {\bibfnamefont {J.}~\bibnamefont {Hoffrogge}},
  \bibinfo {author} {\bibfnamefont {J.}~\bibnamefont {Reichel}}, \bibinfo
  {author} {\bibfnamefont {T.~W.}\ \bibnamefont {H{\"a}nsch}},\ and\ \bibinfo
  {author} {\bibfnamefont {P.}~\bibnamefont {Treutlein}},\ }\bibfield  {title}
  {\bibinfo {title} {Coherent manipulation of bose--einstein condensates with
  state-dependent microwave potentials on an atom chip},\ }\href@noop {}
  {\bibfield  {journal} {\bibinfo  {journal} {Nat. Phys.}\ }\textbf {\bibinfo
  {volume} {5}},\ \bibinfo {pages} {592} (\bibinfo {year} {2009})}\BibitemShut
  {NoStop}%
\bibitem [{\citenamefont {Esteve}\ \emph {et~al.}(2008)\citenamefont {Esteve},
  \citenamefont {Gross}, \citenamefont {Weller}, \citenamefont {Giovanazzi},\
  and\ \citenamefont {Oberthaler}}]{esteve2008squeezing}%
  \BibitemOpen
  \bibfield  {author} {\bibinfo {author} {\bibfnamefont {J.}~\bibnamefont
  {Esteve}}, \bibinfo {author} {\bibfnamefont {C.}~\bibnamefont {Gross}},
  \bibinfo {author} {\bibfnamefont {A.}~\bibnamefont {Weller}}, \bibinfo
  {author} {\bibfnamefont {S.}~\bibnamefont {Giovanazzi}},\ and\ \bibinfo
  {author} {\bibfnamefont {M.}~\bibnamefont {Oberthaler}},\ }\bibfield  {title}
  {\bibinfo {title} {{Squeezing and entanglement in a Bose--Einstein
  condensate}},\ }\href@noop {} {\bibfield  {journal} {\bibinfo  {journal}
  {Nature}\ }\textbf {\bibinfo {volume} {455}},\ \bibinfo {pages} {1216}
  (\bibinfo {year} {2008})}\BibitemShut {NoStop}%
\bibitem [{\citenamefont {Huang}\ and\ \citenamefont
  {Moore}(2008)}]{huang2008optimized}%
  \BibitemOpen
  \bibfield  {author} {\bibinfo {author} {\bibfnamefont {Y.}~\bibnamefont
  {Huang}}\ and\ \bibinfo {author} {\bibfnamefont {M.}~\bibnamefont {Moore}},\
  }\bibfield  {title} {\bibinfo {title} {Optimized double-well quantum
  interferometry with gaussian squeezed states},\ }\href@noop {} {\bibfield
  {journal} {\bibinfo  {journal} {Phys. Rev. Lett.}\ }\textbf {\bibinfo
  {volume} {100}},\ \bibinfo {pages} {250406} (\bibinfo {year}
  {2008})}\BibitemShut {NoStop}%
\bibitem [{\citenamefont {Leroux}\ \emph {et~al.}(2010)\citenamefont {Leroux},
  \citenamefont {Schleier-Smith},\ and\ \citenamefont
  {Vuleti{\'c}}}]{leroux2010orientation}%
  \BibitemOpen
  \bibfield  {author} {\bibinfo {author} {\bibfnamefont {I.~D.}\ \bibnamefont
  {Leroux}}, \bibinfo {author} {\bibfnamefont {M.~H.}\ \bibnamefont
  {Schleier-Smith}},\ and\ \bibinfo {author} {\bibfnamefont {V.}~\bibnamefont
  {Vuleti{\'c}}},\ }\bibfield  {title} {\bibinfo {title} {Orientation-dependent
  entanglement lifetime in a squeezed atomic clock},\ }\href@noop {} {\bibfield
   {journal} {\bibinfo  {journal} {Phys. Rev. Lett.}\ }\textbf {\bibinfo
  {volume} {104}},\ \bibinfo {pages} {250801} (\bibinfo {year}
  {2010})}\BibitemShut {NoStop}%
\bibitem [{\citenamefont {Chen}\ \emph {et~al.}(2011)\citenamefont {Chen},
  \citenamefont {Bohnet}, \citenamefont {Sankar}, \citenamefont {Dai},\ and\
  \citenamefont {Thompson}}]{chen2011conditional}%
  \BibitemOpen
  \bibfield  {author} {\bibinfo {author} {\bibfnamefont {Z.}~\bibnamefont
  {Chen}}, \bibinfo {author} {\bibfnamefont {J.~G.}\ \bibnamefont {Bohnet}},
  \bibinfo {author} {\bibfnamefont {S.~R.}\ \bibnamefont {Sankar}}, \bibinfo
  {author} {\bibfnamefont {J.}~\bibnamefont {Dai}},\ and\ \bibinfo {author}
  {\bibfnamefont {J.~K.}\ \bibnamefont {Thompson}},\ }\bibfield  {title}
  {\bibinfo {title} {Conditional spin squeezing of a large ensemble via the
  vacuum rabi splitting},\ }\href@noop {} {\bibfield  {journal} {\bibinfo
  {journal} {Phys. Rev. Lett.}\ }\textbf {\bibinfo {volume} {106}},\ \bibinfo
  {pages} {133601} (\bibinfo {year} {2011})}\BibitemShut {NoStop}%
\bibitem [{\citenamefont {Berrada}\ \emph {et~al.}(2013)\citenamefont
  {Berrada}, \citenamefont {van Frank}, \citenamefont {B{\"u}cker},
  \citenamefont {Schumm}, \citenamefont {Schaff},\ and\ \citenamefont
  {Schmiedmayer}}]{berrada2013integrated}%
  \BibitemOpen
  \bibfield  {author} {\bibinfo {author} {\bibfnamefont {T.}~\bibnamefont
  {Berrada}}, \bibinfo {author} {\bibfnamefont {S.}~\bibnamefont {van Frank}},
  \bibinfo {author} {\bibfnamefont {R.}~\bibnamefont {B{\"u}cker}}, \bibinfo
  {author} {\bibfnamefont {T.}~\bibnamefont {Schumm}}, \bibinfo {author}
  {\bibfnamefont {J.-F.}\ \bibnamefont {Schaff}},\ and\ \bibinfo {author}
  {\bibfnamefont {J.}~\bibnamefont {Schmiedmayer}},\ }\bibfield  {title}
  {\bibinfo {title} {Integrated mach--zehnder interferometer for bose--einstein
  condensates},\ }\href@noop {} {\bibfield  {journal} {\bibinfo  {journal}
  {Nat. Commun.}\ }\textbf {\bibinfo {volume} {4}} (\bibinfo {year}
  {2013})}\BibitemShut {NoStop}%
\bibitem [{\citenamefont {B\"ucker}\ \emph {et~al.}(2011)\citenamefont
  {B\"ucker}, \citenamefont {Grond}, \citenamefont {Manz}, \citenamefont
  {Berrada}, \citenamefont {Betz}, \citenamefont {Koller}, \citenamefont
  {Hohenester}, \citenamefont {Schumm}, \citenamefont {Perrin},\ and\
  \citenamefont {Schmiedmayer}}]{twin_beam}%
  \BibitemOpen
  \bibfield  {author} {\bibinfo {author} {\bibfnamefont {R.}~\bibnamefont
  {B\"ucker}}, \bibinfo {author} {\bibfnamefont {J.}~\bibnamefont {Grond}},
  \bibinfo {author} {\bibfnamefont {S.}~\bibnamefont {Manz}}, \bibinfo {author}
  {\bibfnamefont {T.}~\bibnamefont {Berrada}}, \bibinfo {author} {\bibfnamefont
  {T.}~\bibnamefont {Betz}}, \bibinfo {author} {\bibfnamefont {C.}~\bibnamefont
  {Koller}}, \bibinfo {author} {\bibfnamefont {U.}~\bibnamefont {Hohenester}},
  \bibinfo {author} {\bibfnamefont {T.}~\bibnamefont {Schumm}}, \bibinfo
  {author} {\bibfnamefont {A.}~\bibnamefont {Perrin}},\ and\ \bibinfo {author}
  {\bibfnamefont {J.}~\bibnamefont {Schmiedmayer}},\ }\bibfield  {title}
  {\bibinfo {title} {Twin-atom beams},\ }\href@noop {} {\bibfield  {journal}
  {\bibinfo  {journal} {Nat. Phys.}\ }\textbf {\bibinfo {volume} {7}},\
  \bibinfo {pages} {608} (\bibinfo {year} {2011})}\BibitemShut {NoStop}%
\bibitem [{\citenamefont {Kheruntsyan}\ \emph {et~al.}(2012)\citenamefont
  {Kheruntsyan}, \citenamefont {Jaskula}, \citenamefont {Deuar}, \citenamefont
  {Bonneau}, \citenamefont {Partridge}, \citenamefont {Ruaudel}, \citenamefont
  {Lopes}, \citenamefont {Boiron},\ and\ \citenamefont
  {Westbrook}}]{cauchy_paris}%
  \BibitemOpen
  \bibfield  {author} {\bibinfo {author} {\bibfnamefont {K.~V.}\ \bibnamefont
  {Kheruntsyan}}, \bibinfo {author} {\bibfnamefont {J.-C.}\ \bibnamefont
  {Jaskula}}, \bibinfo {author} {\bibfnamefont {P.}~\bibnamefont {Deuar}},
  \bibinfo {author} {\bibfnamefont {M.}~\bibnamefont {Bonneau}}, \bibinfo
  {author} {\bibfnamefont {G.~B.}\ \bibnamefont {Partridge}}, \bibinfo {author}
  {\bibfnamefont {J.}~\bibnamefont {Ruaudel}}, \bibinfo {author} {\bibfnamefont
  {R.}~\bibnamefont {Lopes}}, \bibinfo {author} {\bibfnamefont
  {D.}~\bibnamefont {Boiron}},\ and\ \bibinfo {author} {\bibfnamefont {C.~I.}\
  \bibnamefont {Westbrook}},\ }\bibfield  {title} {\bibinfo {title} {Violation
  of the cauchy-schwarz inequality with matter waves},\ }\href@noop {}
  {\bibfield  {journal} {\bibinfo  {journal} {Phys. Rev. Lett.}\ }\textbf
  {\bibinfo {volume} {108}},\ \bibinfo {pages} {260401} (\bibinfo {year}
  {2012})}\BibitemShut {NoStop}%
\bibitem [{\citenamefont {Perrin}\ \emph {et~al.}(2007)\citenamefont {Perrin},
  \citenamefont {Chang}, \citenamefont {Krachmalnicoff}, \citenamefont
  {Schellekens}, \citenamefont {Boiron}, \citenamefont {Aspect},\ and\
  \citenamefont {Westbrook}}]{collision_paris}%
  \BibitemOpen
  \bibfield  {author} {\bibinfo {author} {\bibfnamefont {A.}~\bibnamefont
  {Perrin}}, \bibinfo {author} {\bibfnamefont {H.}~\bibnamefont {Chang}},
  \bibinfo {author} {\bibfnamefont {V.}~\bibnamefont {Krachmalnicoff}},
  \bibinfo {author} {\bibfnamefont {M.}~\bibnamefont {Schellekens}}, \bibinfo
  {author} {\bibfnamefont {D.}~\bibnamefont {Boiron}}, \bibinfo {author}
  {\bibfnamefont {A.}~\bibnamefont {Aspect}},\ and\ \bibinfo {author}
  {\bibfnamefont {C.~I.}\ \bibnamefont {Westbrook}},\ }\bibfield  {title}
  {\bibinfo {title} {Observation of atom pairs in spontaneous four-wave mixing
  of two colliding bose-einstein condensates},\ }\href@noop {} {\bibfield
  {journal} {\bibinfo  {journal} {Phys. Rev. Lett.}\ }\textbf {\bibinfo
  {volume} {99}},\ \bibinfo {pages} {150405} (\bibinfo {year}
  {2007})}\BibitemShut {NoStop}%
\bibitem [{\citenamefont {Bonneau}\ \emph {et~al.}(2013)\citenamefont
  {Bonneau}, \citenamefont {Ruaudel}, \citenamefont {Lopes}, \citenamefont
  {Jaskula}, \citenamefont {Aspect}, \citenamefont {Boiron},\ and\
  \citenamefont {Westbrook}}]{twin_paris}%
  \BibitemOpen
  \bibfield  {author} {\bibinfo {author} {\bibfnamefont {M.}~\bibnamefont
  {Bonneau}}, \bibinfo {author} {\bibfnamefont {J.}~\bibnamefont {Ruaudel}},
  \bibinfo {author} {\bibfnamefont {R.}~\bibnamefont {Lopes}}, \bibinfo
  {author} {\bibfnamefont {J.-C.}\ \bibnamefont {Jaskula}}, \bibinfo {author}
  {\bibfnamefont {A.}~\bibnamefont {Aspect}}, \bibinfo {author} {\bibfnamefont
  {D.}~\bibnamefont {Boiron}},\ and\ \bibinfo {author} {\bibfnamefont {C.~I.}\
  \bibnamefont {Westbrook}},\ }\bibfield  {title} {\bibinfo {title} {Tunable
  source of correlated atom beams},\ }\href@noop {} {\bibfield  {journal}
  {\bibinfo  {journal} {Phys. Rev. A}\ }\textbf {\bibinfo {volume} {87}},\
  \bibinfo {pages} {061603} (\bibinfo {year} {2013})}\BibitemShut {NoStop}%
\bibitem [{\citenamefont {Vogels}\ \emph {et~al.}(2002)\citenamefont {Vogels},
  \citenamefont {Xu},\ and\ \citenamefont {Ketterle}}]{ketterle}%
  \BibitemOpen
  \bibfield  {author} {\bibinfo {author} {\bibfnamefont {J.~M.}\ \bibnamefont
  {Vogels}}, \bibinfo {author} {\bibfnamefont {K.}~\bibnamefont {Xu}},\ and\
  \bibinfo {author} {\bibfnamefont {W.}~\bibnamefont {Ketterle}},\ }\bibfield
  {title} {\bibinfo {title} {Generation of macroscopic pair-correlated atomic
  beams by four-wave mixing in bose-einstein condensates},\ }\href@noop {}
  {\bibfield  {journal} {\bibinfo  {journal} {Phys. Rev. Lett.}\ }\textbf
  {\bibinfo {volume} {89}},\ \bibinfo {pages} {020401} (\bibinfo {year}
  {2002})}\BibitemShut {NoStop}%
\bibitem [{\citenamefont {RuGway}\ \emph {et~al.}(2011)\citenamefont {RuGway},
  \citenamefont {Hodgman}, \citenamefont {Dall}, \citenamefont {Johnsson},\
  and\ \citenamefont {Truscott}}]{truscott}%
  \BibitemOpen
  \bibfield  {author} {\bibinfo {author} {\bibfnamefont {W.}~\bibnamefont
  {RuGway}}, \bibinfo {author} {\bibfnamefont {S.~S.}\ \bibnamefont {Hodgman}},
  \bibinfo {author} {\bibfnamefont {R.~G.}\ \bibnamefont {Dall}}, \bibinfo
  {author} {\bibfnamefont {M.~T.}\ \bibnamefont {Johnsson}},\ and\ \bibinfo
  {author} {\bibfnamefont {A.~G.}\ \bibnamefont {Truscott}},\ }\bibfield
  {title} {\bibinfo {title} {Correlations in amplified four-wave mixing of
  matter waves},\ }\href@noop {} {\bibfield  {journal} {\bibinfo  {journal}
  {Phys. Rev. Lett.}\ }\textbf {\bibinfo {volume} {107}},\ \bibinfo {pages}
  {075301} (\bibinfo {year} {2011})}\BibitemShut {NoStop}%
\bibitem [{\citenamefont {Shin}\ \emph {et~al.}(2019)\citenamefont {Shin},
  \citenamefont {Henson}, \citenamefont {Hodgman}, \citenamefont {Wasak},
  \citenamefont {Chwede\'nczuk},\ and\ \citenamefont
  {Truscott}}]{shin2019bell}%
  \BibitemOpen
  \bibfield  {author} {\bibinfo {author} {\bibfnamefont {D.~K.}\ \bibnamefont
  {Shin}}, \bibinfo {author} {\bibfnamefont {B.~M.}\ \bibnamefont {Henson}},
  \bibinfo {author} {\bibfnamefont {S.~S.}\ \bibnamefont {Hodgman}}, \bibinfo
  {author} {\bibfnamefont {T.}~\bibnamefont {Wasak}}, \bibinfo {author}
  {\bibfnamefont {J.}~\bibnamefont {Chwede\'nczuk}},\ and\ \bibinfo {author}
  {\bibfnamefont {A.~G.}\ \bibnamefont {Truscott}},\ }\bibfield  {title}
  {\bibinfo {title} {Bell correlations between spatially separated pairs of
  atom},\ }\href@noop {} {\bibfield  {journal} {\bibinfo  {journal} {Nat.
  Comm.}\ }\textbf {\bibinfo {volume} {10}},\ \bibinfo {pages} {4447} (\bibinfo
  {year} {2019})}\BibitemShut {NoStop}%
\bibitem [{\citenamefont {L\"ucke}\ \emph {et~al.}(2014)\citenamefont
  {L\"ucke}, \citenamefont {Peise}, \citenamefont {Vitagliano}, \citenamefont
  {Arlt}, \citenamefont {Santos}, \citenamefont {T\'oth},\ and\ \citenamefont
  {Klempt}}]{PhysRevLett.112.155304}%
  \BibitemOpen
  \bibfield  {author} {\bibinfo {author} {\bibfnamefont {B.}~\bibnamefont
  {L\"ucke}}, \bibinfo {author} {\bibfnamefont {J.}~\bibnamefont {Peise}},
  \bibinfo {author} {\bibfnamefont {G.}~\bibnamefont {Vitagliano}}, \bibinfo
  {author} {\bibfnamefont {J.}~\bibnamefont {Arlt}}, \bibinfo {author}
  {\bibfnamefont {L.}~\bibnamefont {Santos}}, \bibinfo {author} {\bibfnamefont
  {G.}~\bibnamefont {T\'oth}},\ and\ \bibinfo {author} {\bibfnamefont
  {C.}~\bibnamefont {Klempt}},\ }\bibfield  {title} {\bibinfo {title}
  {Detecting multiparticle entanglement of dicke states},\ }\href
  {https://doi.org/10.1103/PhysRevLett.112.155304} {\bibfield  {journal}
  {\bibinfo  {journal} {Phys. Rev. Lett.}\ }\textbf {\bibinfo {volume} {112}},\
  \bibinfo {pages} {155304} (\bibinfo {year} {2014})}\BibitemShut {NoStop}%
\bibitem [{\citenamefont {Lücke}\ \emph {et~al.}(2011)\citenamefont {Lücke},
  \citenamefont {Scherer}, \citenamefont {Kruse}, \citenamefont {Pezzé},
  \citenamefont {Deuretzbacher}, \citenamefont {Hyllus}, \citenamefont {Topic},
  \citenamefont {Peise}, \citenamefont {Ertmer}, \citenamefont {Arlt},
  \citenamefont {Santos}, \citenamefont {Smerzi},\ and\ \citenamefont
  {Klempt}}]{doi:10.1126/science.1208798}%
  \BibitemOpen
  \bibfield  {author} {\bibinfo {author} {\bibfnamefont {B.}~\bibnamefont
  {Lücke}}, \bibinfo {author} {\bibfnamefont {M.}~\bibnamefont {Scherer}},
  \bibinfo {author} {\bibfnamefont {J.}~\bibnamefont {Kruse}}, \bibinfo
  {author} {\bibfnamefont {L.}~\bibnamefont {Pezzé}}, \bibinfo {author}
  {\bibfnamefont {F.}~\bibnamefont {Deuretzbacher}}, \bibinfo {author}
  {\bibfnamefont {P.}~\bibnamefont {Hyllus}}, \bibinfo {author} {\bibfnamefont
  {O.}~\bibnamefont {Topic}}, \bibinfo {author} {\bibfnamefont
  {J.}~\bibnamefont {Peise}}, \bibinfo {author} {\bibfnamefont
  {W.}~\bibnamefont {Ertmer}}, \bibinfo {author} {\bibfnamefont
  {J.}~\bibnamefont {Arlt}}, \bibinfo {author} {\bibfnamefont {L.}~\bibnamefont
  {Santos}}, \bibinfo {author} {\bibfnamefont {A.}~\bibnamefont {Smerzi}},\
  and\ \bibinfo {author} {\bibfnamefont {C.}~\bibnamefont {Klempt}},\
  }\bibfield  {title} {\bibinfo {title} {Twin matter waves for interferometry
  beyond the classical limit},\ }\href
  {https://doi.org/10.1126/science.1208798} {\bibfield  {journal} {\bibinfo
  {journal} {Science}\ }\textbf {\bibinfo {volume} {334}},\ \bibinfo {pages}
  {773} (\bibinfo {year} {2011})},\ \Eprint
  {https://arxiv.org/abs/https://www.science.org/doi/pdf/10.1126/science.1208798}
  {https://www.science.org/doi/pdf/10.1126/science.1208798} \BibitemShut
  {NoStop}%
\bibitem [{\citenamefont {Burnham}\ and\ \citenamefont
  {Weinberg}(1970)}]{pdc1}%
  \BibitemOpen
  \bibfield  {author} {\bibinfo {author} {\bibfnamefont {D.~C.}\ \bibnamefont
  {Burnham}}\ and\ \bibinfo {author} {\bibfnamefont {D.~L.}\ \bibnamefont
  {Weinberg}},\ }\bibfield  {title} {\bibinfo {title} {Observation of
  simultaneity in parametric production of optical photon pairs},\ }\href@noop
  {} {\bibfield  {journal} {\bibinfo  {journal} {Phys. Rev. Lett.}\ }\textbf
  {\bibinfo {volume} {25}},\ \bibinfo {pages} {84} (\bibinfo {year}
  {1970})}\BibitemShut {NoStop}%
\bibitem [{\citenamefont {Kwiat}\ \emph {et~al.}(1995)\citenamefont {Kwiat},
  \citenamefont {Mattle}, \citenamefont {Weinfurter}, \citenamefont
  {Zeilinger}, \citenamefont {Sergienko},\ and\ \citenamefont {Shih}}]{pdc2}%
  \BibitemOpen
  \bibfield  {author} {\bibinfo {author} {\bibfnamefont {P.~G.}\ \bibnamefont
  {Kwiat}}, \bibinfo {author} {\bibfnamefont {K.}~\bibnamefont {Mattle}},
  \bibinfo {author} {\bibfnamefont {H.}~\bibnamefont {Weinfurter}}, \bibinfo
  {author} {\bibfnamefont {A.}~\bibnamefont {Zeilinger}}, \bibinfo {author}
  {\bibfnamefont {A.~V.}\ \bibnamefont {Sergienko}},\ and\ \bibinfo {author}
  {\bibfnamefont {Y.}~\bibnamefont {Shih}},\ }\bibfield  {title} {\bibinfo
  {title} {New high-intensity source of polarization-entangled photon pairs},\
  }\href@noop {} {\bibfield  {journal} {\bibinfo  {journal} {Phys. Rev. Lett.}\
  }\textbf {\bibinfo {volume} {75}},\ \bibinfo {pages} {4337} (\bibinfo {year}
  {1995})}\BibitemShut {NoStop}%
\bibitem [{\citenamefont {Peters}\ \emph
  {et~al.}(2001{\natexlab{a}})\citenamefont {Peters}, \citenamefont {Chung},\
  and\ \citenamefont {Chu}}]{A_Peters_2001}%
  \BibitemOpen
  \bibfield  {author} {\bibinfo {author} {\bibfnamefont {A.}~\bibnamefont
  {Peters}}, \bibinfo {author} {\bibfnamefont {K.~Y.}\ \bibnamefont {Chung}},\
  and\ \bibinfo {author} {\bibfnamefont {S.}~\bibnamefont {Chu}},\ }\bibfield
  {title} {\bibinfo {title} {High-precision gravity measurements using atom
  interferometry},\ }\href {https://doi.org/10.1088/0026-1394/38/1/4}
  {\bibfield  {journal} {\bibinfo  {journal} {Metrologia}\ }\textbf {\bibinfo
  {volume} {38}},\ \bibinfo {pages} {25} (\bibinfo {year}
  {2001}{\natexlab{a}})}\BibitemShut {NoStop}%
\bibitem [{\citenamefont {Tino}(2021)}]{Tino_2021}%
  \BibitemOpen
  \bibfield  {author} {\bibinfo {author} {\bibfnamefont {G.~M.}\ \bibnamefont
  {Tino}},\ }\bibfield  {title} {\bibinfo {title} {Testing gravity with cold
  atom interferometry: results and prospects},\ }\href
  {https://doi.org/10.1088/2058-9565/abd83e} {\bibfield  {journal} {\bibinfo
  {journal} {Quantum Science and Technology}\ }\textbf {\bibinfo {volume}
  {6}},\ \bibinfo {pages} {024014} (\bibinfo {year} {2021})}\BibitemShut
  {NoStop}%
\bibitem [{\citenamefont {Peters}\ \emph {et~al.}(1999)\citenamefont {Peters},
  \citenamefont {Chung},\ and\ \citenamefont {Chu}}]{peters1999measurement}%
  \BibitemOpen
  \bibfield  {author} {\bibinfo {author} {\bibfnamefont {A.}~\bibnamefont
  {Peters}}, \bibinfo {author} {\bibfnamefont {K.~Y.}\ \bibnamefont {Chung}},\
  and\ \bibinfo {author} {\bibfnamefont {S.}~\bibnamefont {Chu}},\ }\bibfield
  {title} {\bibinfo {title} {Measurement of gravitational acceleration by
  dropping atoms},\ }\href@noop {} {\bibfield  {journal} {\bibinfo  {journal}
  {Nature}\ }\textbf {\bibinfo {volume} {400}},\ \bibinfo {pages} {849}
  (\bibinfo {year} {1999})}\BibitemShut {NoStop}%
\bibitem [{\citenamefont {Peters}\ \emph
  {et~al.}(2001{\natexlab{b}})\citenamefont {Peters}, \citenamefont {Chung},\
  and\ \citenamefont {Chu}}]{peters2001high}%
  \BibitemOpen
  \bibfield  {author} {\bibinfo {author} {\bibfnamefont {A.}~\bibnamefont
  {Peters}}, \bibinfo {author} {\bibfnamefont {K.~Y.}\ \bibnamefont {Chung}},\
  and\ \bibinfo {author} {\bibfnamefont {S.}~\bibnamefont {Chu}},\ }\bibfield
  {title} {\bibinfo {title} {High-precision gravity measurements using atom
  interferometry},\ }\href@noop {} {\bibfield  {journal} {\bibinfo  {journal}
  {Metrologia}\ }\textbf {\bibinfo {volume} {38}},\ \bibinfo {pages} {25}
  (\bibinfo {year} {2001}{\natexlab{b}})}\BibitemShut {NoStop}%
\bibitem [{\citenamefont {Weiss}\ \emph {et~al.}(1994)\citenamefont {Weiss},
  \citenamefont {Young},\ and\ \citenamefont {Chu}}]{weiss1994precision}%
  \BibitemOpen
  \bibfield  {author} {\bibinfo {author} {\bibfnamefont {D.~S.}\ \bibnamefont
  {Weiss}}, \bibinfo {author} {\bibfnamefont {B.~C.}\ \bibnamefont {Young}},\
  and\ \bibinfo {author} {\bibfnamefont {S.}~\bibnamefont {Chu}},\ }\bibfield
  {title} {\bibinfo {title} {Precision measurement of $\hbar/m$ cs based on
  photon recoil using laser-cooled atoms and atomic interferometry},\
  }\href@noop {} {\bibfield  {journal} {\bibinfo  {journal} {Applied physics
  B}\ }\textbf {\bibinfo {volume} {59}},\ \bibinfo {pages} {217} (\bibinfo
  {year} {1994})}\BibitemShut {NoStop}%
\bibitem [{\citenamefont {Altin}\ \emph {et~al.}(2013)\citenamefont {Altin},
  \citenamefont {Johnsson}, \citenamefont {Negnevitsky}, \citenamefont
  {Dennis}, \citenamefont {Anderson}, \citenamefont {Debs}, \citenamefont
  {Szigeti}, \citenamefont {Hardman}, \citenamefont {Bennetts}, \citenamefont
  {McDonald}, \citenamefont {Turner}, \citenamefont {Close},\ and\
  \citenamefont {Robins}}]{Altin_2013}%
  \BibitemOpen
  \bibfield  {author} {\bibinfo {author} {\bibfnamefont {P.~A.}\ \bibnamefont
  {Altin}}, \bibinfo {author} {\bibfnamefont {M.~T.}\ \bibnamefont {Johnsson}},
  \bibinfo {author} {\bibfnamefont {V.}~\bibnamefont {Negnevitsky}}, \bibinfo
  {author} {\bibfnamefont {G.~R.}\ \bibnamefont {Dennis}}, \bibinfo {author}
  {\bibfnamefont {R.~P.}\ \bibnamefont {Anderson}}, \bibinfo {author}
  {\bibfnamefont {J.~E.}\ \bibnamefont {Debs}}, \bibinfo {author}
  {\bibfnamefont {S.~S.}\ \bibnamefont {Szigeti}}, \bibinfo {author}
  {\bibfnamefont {K.~S.}\ \bibnamefont {Hardman}}, \bibinfo {author}
  {\bibfnamefont {S.}~\bibnamefont {Bennetts}}, \bibinfo {author}
  {\bibfnamefont {G.~D.}\ \bibnamefont {McDonald}}, \bibinfo {author}
  {\bibfnamefont {L.~D.}\ \bibnamefont {Turner}}, \bibinfo {author}
  {\bibfnamefont {J.~D.}\ \bibnamefont {Close}},\ and\ \bibinfo {author}
  {\bibfnamefont {N.~P.}\ \bibnamefont {Robins}},\ }\bibfield  {title}
  {\bibinfo {title} {Precision atomic gravimeter based on bragg diffraction},\
  }\href {https://doi.org/10.1088/1367-2630/15/2/023009} {\bibfield  {journal}
  {\bibinfo  {journal} {New Journal of Physics}\ }\textbf {\bibinfo {volume}
  {15}},\ \bibinfo {pages} {023009} (\bibinfo {year} {2013})}\BibitemShut
  {NoStop}%
\bibitem [{\citenamefont {Masi}\ \emph {et~al.}(2021)\citenamefont {Masi},
  \citenamefont {Petrucciani}, \citenamefont {Ferioli}, \citenamefont
  {Semeghini}, \citenamefont {Modugno}, \citenamefont {Inguscio},\ and\
  \citenamefont {Fattori}}]{masi2021spatial}%
  \BibitemOpen
  \bibfield  {author} {\bibinfo {author} {\bibfnamefont {L.}~\bibnamefont
  {Masi}}, \bibinfo {author} {\bibfnamefont {T.}~\bibnamefont {Petrucciani}},
  \bibinfo {author} {\bibfnamefont {G.}~\bibnamefont {Ferioli}}, \bibinfo
  {author} {\bibfnamefont {G.}~\bibnamefont {Semeghini}}, \bibinfo {author}
  {\bibfnamefont {G.}~\bibnamefont {Modugno}}, \bibinfo {author} {\bibfnamefont
  {M.}~\bibnamefont {Inguscio}},\ and\ \bibinfo {author} {\bibfnamefont
  {M.}~\bibnamefont {Fattori}},\ }\bibfield  {title} {\bibinfo {title} {Spatial
  bloch oscillations of a quantum gas in a “beat-note” superlattice},\
  }\href@noop {} {\bibfield  {journal} {\bibinfo  {journal} {Physical Review
  Letters}\ }\textbf {\bibinfo {volume} {127}},\ \bibinfo {pages} {020601}
  (\bibinfo {year} {2021})}\BibitemShut {NoStop}%
\bibitem [{\citenamefont {Petrucciani}\ \emph {et~al.}(2025)\citenamefont
  {Petrucciani}, \citenamefont {Santoni}, \citenamefont {Mazzinghi},
  \citenamefont {Trypogeorgos}, \citenamefont {Cataliotti}, \citenamefont
  {Inguscio}, \citenamefont {Modugno}, \citenamefont {Smerzi}, \citenamefont
  {Pezz{\'e}},\ and\ \citenamefont {Fattori}}]{petrucciani2025mach}%
  \BibitemOpen
  \bibfield  {author} {\bibinfo {author} {\bibfnamefont {T.}~\bibnamefont
  {Petrucciani}}, \bibinfo {author} {\bibfnamefont {A.}~\bibnamefont
  {Santoni}}, \bibinfo {author} {\bibfnamefont {C.}~\bibnamefont {Mazzinghi}},
  \bibinfo {author} {\bibfnamefont {D.}~\bibnamefont {Trypogeorgos}}, \bibinfo
  {author} {\bibfnamefont {F.~S.}\ \bibnamefont {Cataliotti}}, \bibinfo
  {author} {\bibfnamefont {M.}~\bibnamefont {Inguscio}}, \bibinfo {author}
  {\bibfnamefont {G.}~\bibnamefont {Modugno}}, \bibinfo {author} {\bibfnamefont
  {A.}~\bibnamefont {Smerzi}}, \bibinfo {author} {\bibfnamefont
  {L.}~\bibnamefont {Pezz{\'e}}},\ and\ \bibinfo {author} {\bibfnamefont
  {M.}~\bibnamefont {Fattori}},\ }\bibfield  {title} {\bibinfo {title}
  {Mach-zehnder atom interferometry with non-interacting trapped bose einstein
  condensates},\ }\href@noop {} {\bibfield  {journal} {\bibinfo  {journal}
  {arXiv preprint arXiv:2504.17391}\ } (\bibinfo {year} {2025})}\BibitemShut
  {NoStop}%
\bibitem [{\citenamefont {Hong}\ \emph {et~al.}(1987)\citenamefont {Hong},
  \citenamefont {Ou},\ and\ \citenamefont {Mandel}}]{hong}%
  \BibitemOpen
  \bibfield  {author} {\bibinfo {author} {\bibfnamefont {C.~K.}\ \bibnamefont
  {Hong}}, \bibinfo {author} {\bibfnamefont {Z.~Y.}\ \bibnamefont {Ou}},\ and\
  \bibinfo {author} {\bibfnamefont {L.}~\bibnamefont {Mandel}},\ }\bibfield
  {title} {\bibinfo {title} {Measurement of subpicosecond time intervals
  between two photons by interference},\ }\href@noop {} {\bibfield  {journal}
  {\bibinfo  {journal} {Phys. Rev. Lett.}\ }\textbf {\bibinfo {volume} {59}},\
  \bibinfo {pages} {2044} (\bibinfo {year} {1987})}\BibitemShut {NoStop}%
\bibitem [{\citenamefont {P{\l}odzie{\'n}}\ \emph {et~al.}(2022)\citenamefont
  {P{\l}odzie{\'n}}, \citenamefont {Lewenstein}, \citenamefont {Witkowska},\
  and\ \citenamefont {Chwede{\'n}czuk}}]{plodzien2022one}%
  \BibitemOpen
  \bibfield  {author} {\bibinfo {author} {\bibfnamefont {M.}~\bibnamefont
  {P{\l}odzie{\'n}}}, \bibinfo {author} {\bibfnamefont {M.}~\bibnamefont
  {Lewenstein}}, \bibinfo {author} {\bibfnamefont {E.}~\bibnamefont
  {Witkowska}},\ and\ \bibinfo {author} {\bibfnamefont {J.}~\bibnamefont
  {Chwede{\'n}czuk}},\ }\bibfield  {title} {\bibinfo {title} {One-axis twisting
  as a method of generating many-body bell correlations},\ }\href@noop {}
  {\bibfield  {journal} {\bibinfo  {journal} {Physical Review Letters}\
  }\textbf {\bibinfo {volume} {129}},\ \bibinfo {pages} {250402} (\bibinfo
  {year} {2022})}\BibitemShut {NoStop}%
\bibitem [{\citenamefont {Gerry}\ and\ \citenamefont
  {Knight}(2004)}]{knight_qo}%
  \BibitemOpen
  \bibfield  {author} {\bibinfo {author} {\bibfnamefont {C.}~\bibnamefont
  {Gerry}}\ and\ \bibinfo {author} {\bibfnamefont {P.}~\bibnamefont {Knight}},\
  }\href@noop {} {\emph {\bibinfo {title} {Introductory Quantum Optics}}}\
  (\bibinfo  {publisher} {Cambridge University Press},\ \bibinfo {year}
  {2004})\BibitemShut {NoStop}%
\bibitem [{\citenamefont {Lopes}\ \emph {et~al.}(2015)\citenamefont {Lopes},
  \citenamefont {Imanaliev}, \citenamefont {Aspect}, \citenamefont {Cheneau},
  \citenamefont {Boiron},\ and\ \citenamefont {Westbrook}}]{lopes2015atomic}%
  \BibitemOpen
  \bibfield  {author} {\bibinfo {author} {\bibfnamefont {R.}~\bibnamefont
  {Lopes}}, \bibinfo {author} {\bibfnamefont {A.}~\bibnamefont {Imanaliev}},
  \bibinfo {author} {\bibfnamefont {A.}~\bibnamefont {Aspect}}, \bibinfo
  {author} {\bibfnamefont {M.}~\bibnamefont {Cheneau}}, \bibinfo {author}
  {\bibfnamefont {D.}~\bibnamefont {Boiron}},\ and\ \bibinfo {author}
  {\bibfnamefont {C.~I.}\ \bibnamefont {Westbrook}},\ }\bibfield  {title}
  {\bibinfo {title} {Atomic hong-ou-mandel experiment},\ }\href@noop {}
  {\bibfield  {journal} {\bibinfo  {journal} {Nature}\ }\textbf {\bibinfo
  {volume} {520}},\ \bibinfo {pages} {66} (\bibinfo {year} {2015})}\BibitemShut
  {NoStop}%
\bibitem [{\citenamefont {Killoran}\ \emph {et~al.}(2014)\citenamefont
  {Killoran}, \citenamefont {Cramer},\ and\ \citenamefont
  {Plenio}}]{identical_plenio}%
  \BibitemOpen
  \bibfield  {author} {\bibinfo {author} {\bibfnamefont {N.}~\bibnamefont
  {Killoran}}, \bibinfo {author} {\bibfnamefont {M.}~\bibnamefont {Cramer}},\
  and\ \bibinfo {author} {\bibfnamefont {M.~B.}\ \bibnamefont {Plenio}},\
  }\bibfield  {title} {\bibinfo {title} {Extracting entanglement from identical
  particles},\ }\href@noop {} {\bibfield  {journal} {\bibinfo  {journal} {Phys.
  Rev. Lett.}\ }\textbf {\bibinfo {volume} {112}},\ \bibinfo {pages} {150501}
  (\bibinfo {year} {2014})}\BibitemShut {NoStop}%
\bibitem [{\citenamefont {Beugnon}\ \emph {et~al.}(2006)\citenamefont
  {Beugnon}, \citenamefont {Jones}, \citenamefont {Dingjan}, \citenamefont
  {Darqui\'e}, \citenamefont {Messin}, \citenamefont {Browaeys},\ and\
  \citenamefont {Grangier}}]{grangier}%
  \BibitemOpen
  \bibfield  {author} {\bibinfo {author} {\bibfnamefont {J.}~\bibnamefont
  {Beugnon}}, \bibinfo {author} {\bibfnamefont {M.~P.~A.}\ \bibnamefont
  {Jones}}, \bibinfo {author} {\bibfnamefont {J.}~\bibnamefont {Dingjan}},
  \bibinfo {author} {\bibfnamefont {B.}~\bibnamefont {Darqui\'e}}, \bibinfo
  {author} {\bibfnamefont {G.}~\bibnamefont {Messin}}, \bibinfo {author}
  {\bibfnamefont {A.}~\bibnamefont {Browaeys}},\ and\ \bibinfo {author}
  {\bibfnamefont {P.}~\bibnamefont {Grangier}},\ }\bibfield  {title} {\bibinfo
  {title} {Quantum interference between two single photons emitted by
  independently trapped atoms},\ }\href@noop {} {\bibfield  {journal} {\bibinfo
   {journal} {Nature}\ }\textbf {\bibinfo {volume} {440}},\ \bibinfo {pages}
  {779} (\bibinfo {year} {2006})}\BibitemShut {NoStop}%
\bibitem [{\citenamefont {Yurke}\ and\ \citenamefont {Stoler}(1992)}]{yurke}%
  \BibitemOpen
  \bibfield  {author} {\bibinfo {author} {\bibfnamefont {B.}~\bibnamefont
  {Yurke}}\ and\ \bibinfo {author} {\bibfnamefont {D.}~\bibnamefont {Stoler}},\
  }\bibfield  {title} {\bibinfo {title} {Bell's-inequality experiments using
  independent-particle sources},\ }\href@noop {} {\bibfield  {journal}
  {\bibinfo  {journal} {Phys. Rev. A}\ }\textbf {\bibinfo {volume} {46}},\
  \bibinfo {pages} {2229} (\bibinfo {year} {1992})}\BibitemShut {NoStop}%
\bibitem [{\citenamefont {Wasak}\ \emph {et~al.}(2015)\citenamefont {Wasak},
  \citenamefont {Sza\ifmmode~\acute{n}\else \'{n}\fi{}kowski},\ and\
  \citenamefont {Chwede\ifmmode~\acute{n}\else
  \'{n}\fi{}czuk}}]{PhysRevA.91.043619}%
  \BibitemOpen
  \bibfield  {author} {\bibinfo {author} {\bibfnamefont {T.}~\bibnamefont
  {Wasak}}, \bibinfo {author} {\bibfnamefont {P.}~\bibnamefont
  {Sza\ifmmode~\acute{n}\else \'{n}\fi{}kowski}},\ and\ \bibinfo {author}
  {\bibfnamefont {J.}~\bibnamefont {Chwede\ifmmode~\acute{n}\else
  \'{n}\fi{}czuk}},\ }\bibfield  {title} {\bibinfo {title} {{Interferometry
  with independently prepared Bose-Einstein condensates}},\ }\href
  {https://doi.org/10.1103/PhysRevA.91.043619} {\bibfield  {journal} {\bibinfo
  {journal} {Phys. Rev. A}\ }\textbf {\bibinfo {volume} {91}},\ \bibinfo
  {pages} {043619} (\bibinfo {year} {2015})}\BibitemShut {NoStop}%
\end{thebibliography}
\end{document}